\documentclass[12pt]{article}
\usepackage{styles/mystyle} 

\begin{document}

\title{Cyber risk and the cross-section of stock returns}

\date{\today}

\author{Daniel Celeny\thanks{Swiss Finance Institute, EPFL - Cyber-Defence Campus, armasuisse S+T. \href{mailto:daniel.celeny@alumni.epfl.ch}{daniel.celeny@alumni.epfl.ch}} \hspace{0.1cm} and 
\hspace{0.1cm} Loïc Maréchal\thanks{HEC Lausanne, University of Lausanne, Cyber-Defence Campus, armasuisse S+T.\href{mailto:loic.marechal@unil.ch}{loic.marechal@unil.ch}\newline\newline This document results from a research project funded by the Cyber-Defence Campus, armasuisse Science and Technology. Preprints of early-stage versions of this research are available on \href{http://dx.doi.org/10.2139/ssrn.4587993}{SSRN} and \href{
https://doi.org/10.48550/arXiv.2402.04775}{arXiv}. We appreciate helpful comments from seminar participants at the Cyber Alp Retreat 2023. We also thank Pierre Collin-Dufresne and Michel Dubois for their invaluable comments. Our code will be available at \url{https://github.com/technometrics-lab/17-Cyber-risk_and_the_cross-section_of_stock_returns} Corresponding author: Daniel Celeny e-mail: \href{daniel.celeny@alumni.epfl.ch}{daniel.celeny@alumni.epfl.ch}, Cyber-Defence Campus, Innovation Park, EPFL, 1015 Lausanne.}}

\renewcommand{\thefootnote}{\fnsymbol{footnote}}
\maketitle
\vspace{-.2in}

\begin{abstract}
\noindent
We extract firms' cyber risk with a machine learning algorithm measuring the proximity between their disclosures and a dedicated cyber corpus. Our approach outperforms dictionary methods, uses full disclosure and not devoted-only sections, and generates a cyber risk score uncorrelated with other firms' 
characteristics. We find that a portfolio of US-listed stocks in the high cyber risk quantile generates an excess return of 18.72\% p.a. Moreover, a long-short cyber risk portfolio has a significant and positive risk premium of 6.93\% p.a., robust to all factors' benchmarks. Finally, using a Bayesian asset pricing method, we show that our cyber risk factor is the essential feature that allows any multi-factor model to price the cross-section of stock returns.\\ \\

\end{abstract}



\thispagestyle{empty}
\clearpage

\begin{center}
    \textbf{Conflict-of-interest disclosure statement}\bigskip
\end{center}

\noindent
Daniel Celeny:\\
I have no conflict of interests and nothing to disclose \bigskip

\noindent
Loïc Maréchal:\\
I have no conflict of interests and nothing to disclose \bigskip

\clearpage

\onehalfspacing
\setcounter{footnote}{0}
\renewcommand{\thefootnote}{\arabic{footnote}}

\section{Introduction}

In 2023, the SEC introduced new regulations impacting cybersecurity compliance for US public companies. These rules, announced by SEC Chair Gary Gensler on July 26, 2023, require disclosure of \quotes{material} cybersecurity incidents within four business days, with exceptions for national security or public safety reasons. Additionally, companies must submit detailed annual reports on cybersecurity risk management, strategy, and governance practices. For annual reports such as Form 10-K, disclosures will be due for fiscal years ending on or after December 15, 2023.\footnote{\href{https://www.sec.gov/news/press-release/2023-139}{https://www.sec.gov/news/press-release/2023-139}.}

%
%
This regulatory change reflects that cyber threats are becoming ever more widespread and costly; similarly, cyber insurance contracts have become vital for public companies and governments, who must assess cyber risk premia. These insurance contracts, however, need a thorough understanding of the systematic and firm-level cyber risks. In a recent interview\footnote{Available at \href{https://www.ft.com/content/63ea94fa-c6fc-449f-b2b8-ea29cc83637d}{https://www.ft.com/content/63ea94fa-c6fc-449f-b2b8-ea29cc83637d}}, Mario Greco, CEO of Zurich Insurance Group, said that cyberattacks are set to become \quotes{uninsurable} and called on governments to \quotes{set up private-public schemes to handle systemic cyber risks that can't be quantified, similar to those that exist in some jurisdictions for earthquakes or terror attacks}.

In this paper, we develop a method to quantify the cyber risk of a company based on its disclosures and investigate whether this risk is costly to firms in the form of a market risk premium that shows up in their stock returns. To do this, we collect financial fillings, monthly returns, and other firm characteristics for over 7,000 firms listed on US stock markets between January 2007 and December 2022. We use a machine learning algorithm, the \quotes{Paragraph Vector}, in combination with the MITRE ATT\&CK cybersecurity knowledgebase to score each firm's filing based on its cybersecurity content.

We find evidence that our cyber risk does not correlate with firm size, book-to-market, beta, and other standard firms' characteristics known to help price stock returns. At the aggregated level, our cyber risk score shows a monotonic increasing trend, moving from 0.51 to 0.54. In contrast, the cross-sectional distribution of that score is exceptionally narrow (standard deviation of 0.03). We compare our cyber risk score across Fama-French 12 industries and find results aligned with the view that \quotes{Business Equipment} and \quotes{Telephone and Television Transmission} are the riskiest and \quotes{Oil and Gas} and \quotes{Utilities}, the safest.

The cyber risk sorted long-short portfolio, which invests in high cyber risk stocks and shorts low cyber risk stocks, has an average annual excess return of 6.93\% and is statistically significant at the 10 or 5\% level even when controlling for common risk factors. This portfolio performs particularly well before the first mention of a cyber risk premium on SSRN in November 2020 by \cite{FlorackisLoucaMichaelyWeber2023}, with an average annual excess return of 11.88\%. It is statistically significant at the 1\% level. Double sorts confirm that cyber risk captures a variation in stock returns when controlling for other factors.

We use asset pricing tests and find that the cyber risk exposure generates a significant premium after controlling for market beta, book-to-market, size, momentum, operating profitability, and investment aggressiveness (see \citealp{FamaFrench2015}). This performance shows up both in cross-section, with \cite{FamaMacBeth1973} regressions, and time series, with no significant joint alphas in \cite{GibbonsRossShanken1989} tests. Using the Bayesian approach of \cite{BarillasShanken2018}, we show that the optimal subset of factors pricing stock returns always includes our cyber risk factor.

We conduct tests to verify the robustness of our factor. First, we control that our baseline estimate, revised at each new filing, captures the latent cyber risk and not the immediate effect of a cyberattack. To do so, we build a long-run cyber risk score capturing the cumulative cyber risk effect. Our results are virtually unchanged. Second, we control for the possibility that firms in the cybersecurity business have risk occurrences that might positively affect them. We also do not find any differences after that control.

Finally, we compare our model to the dictionary approach used by \cite{FlorackisLoucaMichaelyWeber2023}. While the two estimates are positively correlated, our score performs better, especially for firms assigned with zero cyber risk using the dictionary approach. We also argue that previous methods employed to extract cyber risk scores from financial documents need to be revised for this purpose. First, the dictionary approach may miss the context and does not allow for capturing high-level semantics. Second, given this first issue, it also requires limiting the scope of the analysis to a specific section (Item 1.A in the 10-K case). Third, it requires training based on the actual realization of cyberattacks, thereby raising doubts about whether the effect captured is a \quotes{post-attack recovery}, or indeed a latent cyber risk perceived by investors. We demonstrate it conceptually and from our results while finding strong support for the fact that cyber risk is priced in the cross-section of stock returns.

The remainder of the paper proceeds as follows. Section 2 introduces the existing literature and highlights our contributions. Section 3 presents the data and methods, Section 4 details the results, and Section 5 concludes.

\section{Literature review}
\subsection{Economic implications of cyber risk}

Our study contributes to the literature on the economic implications of cyber risk, the costs of cyberattacks, and corporate measures to mitigate them. For instance,
\cite{AmirLeviLivne2018} compare firms with voluntary disclosures and those withholding information. They find that as the former are associated with a 0.7\% decline in share value, the latter experience a 3.6\% decline, the month during which the attack is discovered. \cite{GordonLoebSohail2010} assess the firm value of voluntary cybersecurity disclosures. They find positive correlations between voluntary disclosure and both firm value and liquidity.

\cite{AndersonBartonBöhmeClaytonEetenLeviMooreSavage2013} perform a systematic study of the costs of cybercrimes by types. They find the typical yearly cost for a citizen to lie between hundreds of USD (for tax and welfare fraud) and a few cents (for \quotes{new crime} such as botnets). Revisited by \cite{AndersonBartonBöhmeClaytonGananGrassoLeviMooreVasek2019}, the updated study shows that payment frauds have doubled over the seven years separating the initial one, along with a significant fall in average costs for citizens. \cite{Bouveret2018W} estimates an average cyberattack loss of USD 97 bn at the country level and a value-at-risk in the USD 147-201 bn range. Closely, \cite{Romanosky2016} finds that the cost distribution is heavily skewed, with an average cost of USD 6 mln and a median cost of USD 170,000.

Much literature on the economic impact of cyber incidents involves event studies. For instance, \cite{GordonLoebZhou2011} find that news about information security breaches significantly negatively affects stock returns, although they also document a downward trend of this effect over time. Similarly, \cite{CampbellGordonLoebZhou2003} uncover a systematic adverse market reaction around cyber incidents but only when it involves confidential information. Similarly, studying data breaches \cite{JohnsonKangLawson2017} find that, on average, stocks decrease by 0.37\% around them, whereas \cite{LendingMinnickSchorno2018} uncover a long-term effect and identify one-year buy-and-hold abnormal returns of -3.5\%. Studying the channels by which these abnormal returns around cyber incidents materialize, \cite{Tosun2021,KamiyaKangKimMilidonisStulz2021} attribute their effects to arise from shocks to firms' reputations. Finally, \cite{AndreadisKalotychouLoucaLundbladMakridis2023} focus on the US municipal bonds market and estimates that a 1\% increase in the number of cyberattacks leads to an increase in yields from 3.7 to 5.9 basis points. Conversely, \cite{JensenPaine2023} find no immediate effect on bond yields of hacked towns. Furthermore, in the subsequent 24 months, they document the municipal bond yields to decline along with an increase in IT spending.

\subsection{Information extraction from natural language}

Our study also contributes to a growing body of literature in finance that uses text-extracted information and recent machine-learning advances to treat texts. First, natural language processing in the financial literature often uses sentiment, \textit{i.e.} a concept that parses the single dimension of positive \textit{vs.} negative content. To alleviate this concern, \cite{LoughranMcDonald2011}, while showing that \quotes{standard} dictionaries can not capture the sentiment in financial documents, develop their negative and positive dictionaries and four others to extract other tones.\footnote{The other tones are: uncertainty, litigious, strong modal, and weak modal.}
Similarly, \cite{JegadeeshWu2013} use positive and negative dictionaries to extract sentiment from 10-K statements that they directly map to market reactions. \cite{Garcia2013} uses ratios of positive and negative words from the New York Times news over a century to predict stock returns in the US, while  \cite{CalomirisMamaysky2019} use a similar method to predict risk and returns in stock markets worldwide.

Dictionary-based methods may also capture more meaning than positive or negative sentiment. For instance, \cite{BodnarukLoughranMcDonald2015} study 10-K statements with a dedicated lexicon identifying constraining words and build a measure uncorrelated with traditional financial constraint measures, yet accurately predicting liquidity events. \cite{AntweilerFrank2004} use a measure of stock disagreement on internet forums and identify that disagreement is positively associated with trading volume. \cite{HassanHollandervLentTahoun2019} use word and bigram counts in quarterly earnings conference calls to capture the political risk of US-listed firms, whereas \cite{SautnerLentVilkovZhang2023} use a similar approach in the same documents to capture climate change risk contexts. Finally, \cite{HobergPhillips2010} uses a measure of product market language proximity between firms in 10-K statements and finds that it predicts the likelihood of a future merger or acquisition between them.

Our study finally borrows from recent advances in the machine-learning treatment of texts.
\cite{LeMikolov2014} develop an unsupervised algorithm called \quotes{Paragraph Vector} that can learn fixed-length vector representations from variable-length pieces of texts, such as sentences, paragraphs, and documents. For example, each piece of text is represented by a dense vector that can be used for text classification and sentiment analysis. The advantage of this algorithm over other methods, such as bag-of-words, is that it learns the semantics of words and sentences. \cite{LauBaldwin2016} perform a rigorous empirical evaluation of this algorithm and provide recommendations on hyper-parameter settings for general-purpose applications. \cite{MendsaikhanHasegawaYamaguchiShimada2019} use the Paragraph Vector algorithm with cybersecurity-specific training data to identify cybersecurity-related texts. Their training data combines online forums, news outlets, and vulnerability databases. Their model identifies cybersecurity texts with 83\% accuracy. Finally, close to our research, \cite{AdosoglouLombardoPardalos2021} construct portfolios based on the semantic differences between two consecutive 10-K statements of each firm, thereby revisiting \cite{CohenMalloyNguyen2020}. They find that cosine similarity is the most effective metric for neural network embeddings, such as the ones obtained using the Paragraph Vector algorithm.

\subsection{Asset pricing with cyber risk}

Last, our study relates to asset pricing research, using firms' cyber risk scores as a portfolio-sorting variable, a body of literature in its infancy. \cite{FlorackisLoucaMichaelyWeber2023} build a text-based cyber risk score using the risk-disclosure section of 10-K statements (\quotes{Item 1.A Risk Factors}). They restrict their analysis to cyber-risk-related sentences from this section, extracted with a match on a list of cyber-related keywords. They consider recently (-1 or -2 years before filing) hacked firms as a training sample and compute the cyber risk exposure of firms. This exposure is computed as the average similarity between the bag-of-words representation (vector of the number of occurrences of each word in their dictionary) of the firm's cybersecurity sentences and the cybersecurity sentences of the training sample. They find that stocks with high exposure have higher returns on average but perform worse in periods of cyber risk. \cite{JamilovReyTahoun2023W} perform a similar analysis using quarterly earnings calls worldwide from 2002 to 2021. They construct a cyber risk score using the frequency of cyber-related keywords in earnings calls and find that cyber risk exposure is priced in stocks and options markets and predicts cyberattacks. \cite{JiangKhannaYangZhou2023} use a cyber dictionary from NIST to count the number of cyber-related words in Item 1.A. Combining it with firm characteristics, they perform a logit ridge regression where the dependent variable is the probability for a firm to experience a cyberattack. Thus, this ex-ante probability also provides a measure of the cyber risk faced by a firm. \cite{LiuMarshXiao2022W} use an intertemporal CAPM where the betas are estimated \textit{vs.} a cybercrime news attention proxy (Thomson-Reuters MarketPsych, TRMI). They find that high cybercrime-beta stocks generate 0.6\% lower return than low cybercrime-beta stocks, in apparent contradiction with studies above. However, they likely capture sensitivities to realized cyberattacks, not latent cyber risk.

We are only aware of these four studies focusing on cyber risk and its implications for asset pricing. As they all use a dictionary approach as a starting point, we argue that this method is unsuitable for estimating firm-level cyber risk exposure for several reasons. First, it leaves many firm-year observations with a cyber risk of zero (71\% of firms in 2007 in the case of \cite{FlorackisLoucaMichaelyWeber2023} and over 98\% of earnings calls in 2007 in the case \cite{JamilovReyTahoun2023W}). Second, missing observations or measurement errors sometimes implies to build a common cyber factor from which firm-level betas are estimated instead of directly estimating it at the firm-level (see \citealp{JiangKhannaYangZhou2023}).\footnote{This was also the case in an earlier version of \cite{JamilovReyTahoun2023W}} Third, this approach only considers word presence in disclosures and not the general cyber risk context. Finally, it sometimes requires calibration with respect to realized cyberattacks (see, \citealp{FlorackisLoucaMichaelyWeber2023,JiangKhannaYangZhou2023}), which casts doubt about whether the estimates proxies for latent cyber risk or cyberattack realization.\footnote{\cite{FlorackisLoucaMichaelyWeber2023} present a robustness test to alleviate this concern, in which they withdraw the firms having experienced cyberattacks from the sample, obtaining less significant results. However, by calibrating their risk score on past cyberattacks, they likely capture the probability of a past cyberattack (missed by the news or the cyberattacks database, PRC) rather than capturing contemporaneous cyber risk exposure.} Thus, their score may very well proxy for a post-cyberattack recovery effect. Our contribution is also to clear these concerns up. Using the Paragraph Vector machine learning algorithm of \cite{LeMikolov2014}, our method is free of cyberattack calibration, focuses on the cybersecurity context, and consistently generates a non-zero score. 


    

\section{Data and methodology}
\subsection{Market data}
\label{section:md}

\begin{figure}[h]
    \noindent\makebox[\textwidth]{%
    \includegraphics[width=\textwidth]{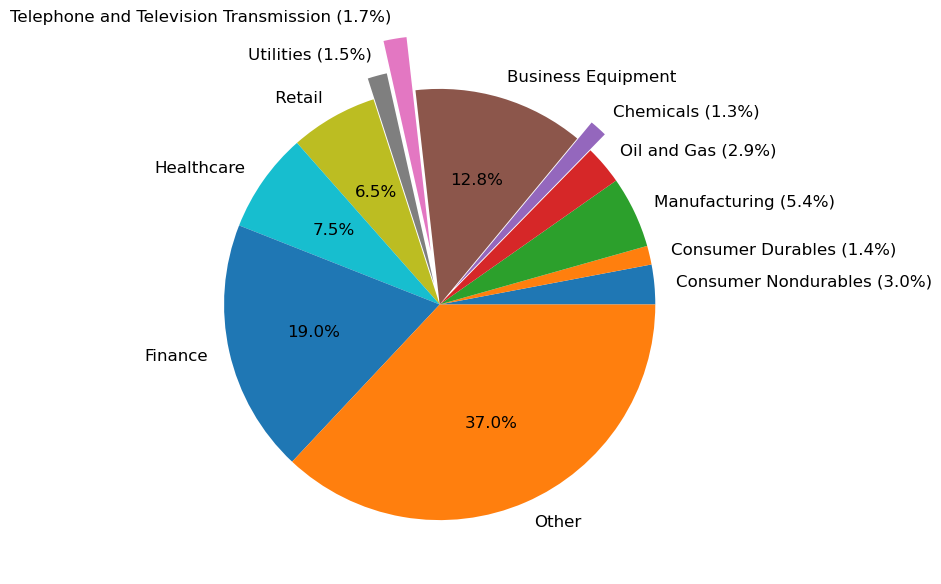}}
     \caption{\textbf{Industry distribution}}\bigskip
     \footnotesize{This figure depicts the distribution of firms in the 12 Fama-French industries Standard Industrial Classification. (SIC) codes are obtained from CRSP. The conversion table, from SIC to 12 Fama-French industries, is available on the \href{https://mba.tuck.dartmouth.edu/pages/faculty/ken.french/Data_Library/det_12_ind_port.html}{Kenneth French data repository}.}
     \label{fig:industry_distributions}
\end{figure}

We download public equity data from Wharton Research Data Services (WRDS), using the data from the Center for Research in Security Prices (CRSP) and S\&P Global Market Intelligence's Compustat database. We report the list of variables in Table \ref{tab:variable_descriptions}. We write a Python script that queries all available information from WRDS' API and filters the firms based on the existence of a 10-K filing with the SEC so that all of the retained firms have at least one 10-K statement available. We extract monthly stock returns and financial ratios for 7,059 firms between January 2007 and December 2022. Figure \ref{fig:industry_distributions} depicts the industry distribution of these firms using the Fama-French 12 industry classification.

We also download the one-month Treasury bill rate and returns on the market, book-to-market (HML), size (SMB), momentum (MOM), investment (CMA), and operating profitability (RMW) factors from the Kenneth French data repository\footnote{Available at: \href{http://mba.tuck.dartmouth.edu/pages/faculty/ken.french/data\_library.html}{http://mba.tuck.dartmouth.edu/pages/faculty/ken.french/data\_library.html}}.
\begin{center}
    [Insert Figure \ref{fig:industry_distributions} here.]
\end{center}

\subsection{10-K statements}
\label{subsection:10K}

10-K statements are financial filings publicly traded companies submit annually to the US Securities and Exchange Commission (SEC). They contain information such as companies' financial statements, risk factors, and executive compensation. We use 10-K statements to build a cyber risk score.

\begin{figure}
    \noindent\makebox[\textwidth]{%
    \includegraphics[width=1\textwidth]{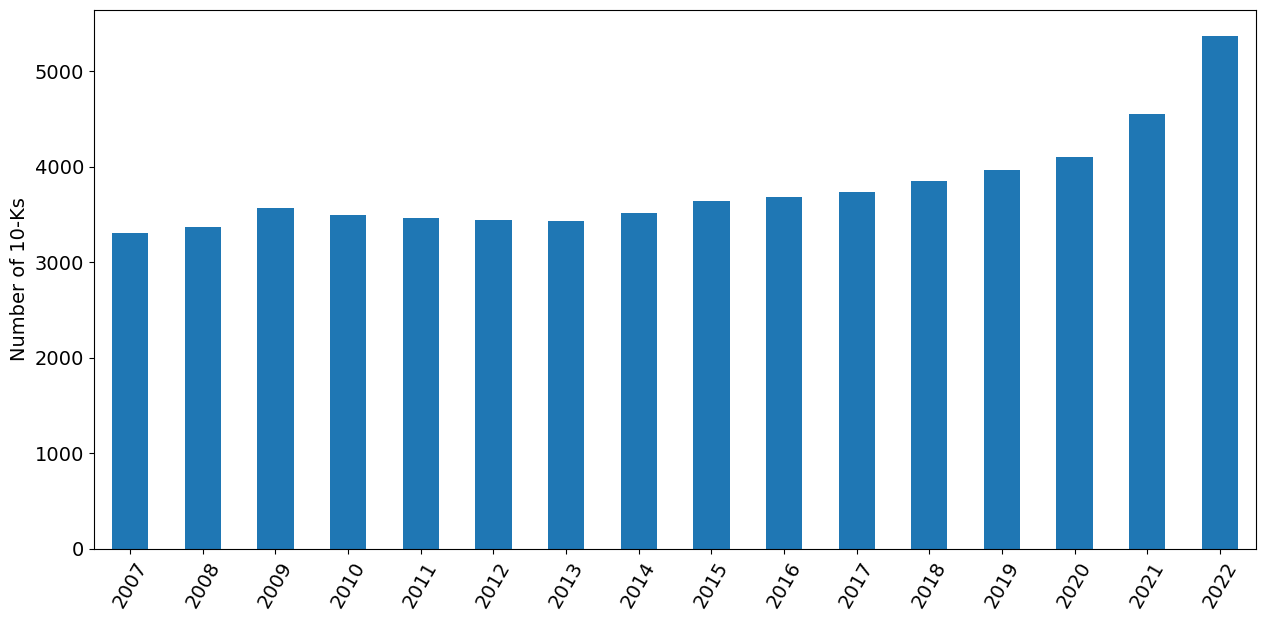}}
     \caption{\textbf{Number of 10-Ks per year}}\bigskip
     \footnotesize{This figure depicts the number of companies in the study sample that have filed a 10-K statement in a given calendar year.}
     \label{fig:number_10Ks}
\end{figure}

To download these statements, we use the index files from the SEC's Edgar archives\footnote{Available at: \href{https://www.sec.gov/Archives/edgar/full-index/}{https://www.sec.gov/Archives/edgar/full-index/}}. These index files contain information about all the documents filed by all firms for a specific quarter. Each line of the index file corresponds to a document and is structured as follows,
\begin{center}
    CIK|Company Name|Form Type|Date Filed|Filename
\end{center}
where Filename is the URL under which an HTML version of the document is available. To identify firms, we use their Central Index Key (CIK), a number used by the SEC to identify corporations and individuals who have filed disclosures. We write a Python script that goes through these index files and identifies URLs that correspond to 10-K statements using the Form Type entry. Using the CIK entry, these URLs are matched to one of the 7,059 firms mentioned in Section \ref{section:md}. We identify 60,470 10-K statements, corresponding to 8.6 statements per firm on average over the sample. Figure \ref{fig:number_10Ks} shows the number of 10-Ks filed per year. This number increases from 3,301 in 2007 to 5,370 in 2022.
\begin{center}
    [Insert Figure \ref{fig:number_10Ks} here.]
\end{center}

\subsection{Cybersecurity tactics}

We use the MITRE ATT\&CK\footnote{Available at: \href{https://attack.mitre.org/}{https://attack.mitre.org/}} cybersecurity knowledge base as a reference for cybersecurity descriptions. This knowledge base was created in 2013 to document cybersecurity tactics, techniques, and procedures used by adversaries against particular platforms, such as Windows or Google Workspace. Figure \ref{fig:MITRE_structure} illustrates the structure of the knowledge base. Each sub-technique has a short description describing it. Table \ref{tab:MITRE_examples} shows two sub-technique descriptions from the knowledge base.
\begin{center}
    [Insert Figure \ref{fig:MITRE_structure} here.]
\end{center}
\begin{center}
    [Insert Table \ref{tab:MITRE_examples} here.]
\end{center}
There are 14 tactics: reconnaissance,  resource development, initial access, execution, persistence, privilege escalation, defense evasion, credential access, discovery, lateral movement, collection, command and control, exfiltration, and impact. There are 785 sub-techniques across all tactics.

\subsection{Methodology}
\label{section:methodology}

\subsubsection{Text preprocessing}
\label{subsection:preprocessing}

We download 10-K statements from the SEC Archives as HTML files. We use the BeautifulSoup\footnote{Available at: \href{https://www.crummy.com/software/BeautifulSoup/}{https://www.crummy.com/software/BeautifulSoup/}} Python library to extract the usable text from these files. We remove the punctuation and numbers and set all letters to lowercase. Given the resulting texts, we write a Python script that uses the wordfreq\footnote{Available at: \href{https://pypi.org/project/wordfreq/}{https://pypi.org/project/wordfreq/}} and NLTK\footnote{Available at: \href{https://www.nltk.org/}{https://www.nltk.org/}} libraries to divide the text into sentences, remove stop-words (\quotes{the}, \quotes{is}, \quotes{and}, \ldots) and remove the most common words of the English language. Since these words frequently appear in the texts, removing them allows us to focus on essential cybersecurity-related words .\footnote{Stop-words contain no valuable information that could hint at the paragraph's content. For this reason, it is standard practice to remove them before using such models (see, \textit{e.g.}, \citealp{MendsaikhanHasegawaYamaguchiShimada2019}). We also attempted to train the models to keep stop-words, but they were then less powerful at identifying cyber-related paragraphs.}

After pre-processing, the average length of the cybersecurity sub-technique descriptions from MITRE ATT\&CK is close to 40 words ($\approx 39.7$). Based on this number, we write a Python algorithm to merge consecutive sentences from 10-K statements into paragraphs with an average length of close to 40 words after pre-processing. On average, we obtain 44 words per paragraph and 638 paragraphs per 10-K statement, with standard deviations of 2.6 words per paragraph and 304 paragraphs per 10-K statement.

\subsubsection{Paragraph Vector}

The Paragraph Vector algorithm, proposed by \cite{LeMikolov2014}, is an extension of the word2vec algorithm (\cite{MikolovChenCorradoDean2013}). It aims to learn fixed-length vector representations from variable-length pieces of text. The main advantage of this algorithm over other methods, such as bag-of-words, is that semantically similar paragraphs are mapped close to each other in the vector space.

\begin{figure}
    \noindent\makebox[1.05\textwidth]{%
    \includegraphics[width=1\textwidth]{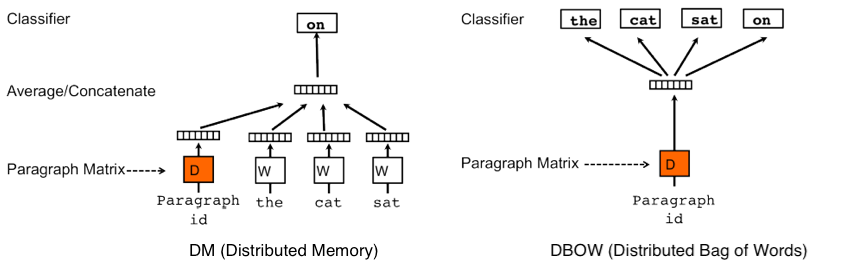}}
     \caption{\textbf{\quotes{Paragraph Vector} model versions}}\bigskip
     \footnotesize{The left panel of the figure depicts the framework for learning paragraph vectors. A paragraph token is mapped to a vector via matrix D (in orange). In this case, the concatenation of this vector with a context of three words is used to predict the fourth word (\quotes{on}). The paragraph vector represents the missing information from the context and is a memory of the paragraph's topic. The right panel depicts the \quotes{Distributed Bag of Words} version of paragraph vectors. This version trains the paragraph vector to predict the words in a small window.
     The images (captions) are taken (rewritten) from \cite{LeMikolov2014}.}
     \label{fig:doc2vec_models}
\end{figure}

The algorithm has two versions: a distributed memory model (DM) and a distributed bag-of-words model (DBOW). In the distributed memory model, the algorithm trains to get both word and paragraph vectors. During training, the concatenation or the average of the paragraph vector and the vector representation of context words are used to predict another word in the paragraph. In the distributed bag-of-words model, the paragraph vector is trained to predict words in a window sampled from the paragraph. Word vectors are not trained in this version. Figure \ref{fig:doc2vec_models} illustrates the two models.
\begin{center}
    [Insert Figure \ref{fig:doc2vec_models} here.]
\end{center}
Both models are unsupervised, as the paragraph vectors are learned from unlabelled data. We use the implementation by Gensim called doc2vec\footnote{Available at: \href{https://radimrehurek.com/gensim/models/doc2vec.html}{https://radimrehurek.com/gensim/models/doc2vec.html}}. To train the model, we use the paragraphs from 10-K statements filed in 2007 as well as the 785 sub-technique descriptions from MITRE, which together amount to more than 1.7 million training paragraphs. Using this sample, we train DM and DBOW doc2vec specifications with various vector dimensions, epochs, and window sizes. The baseline for the hyperparameters is taken from \cite{LauBaldwin2016} (see Table \ref{tab:doc2vec_parameters}).

To choose the best specification, we compute the vector representations of the paragraphs from 10 randomly chosen 10-K statements from 2008 (validation sample) using each specification and compare the highest-scoring paragraphs between the models (the scoring algorithm is explained below). We choose the best specification where the proportion of the highest-scoring cyber risk-related paragraphs is the highest. Table \ref{tab:doc2vec_parameters} presents the parameters of the best-performing doc2vec specification, which is used for the remainder of this study. Appendix Table \ref{tab:top_paragraphs_doc2vec_validation} gives an excerpt of top-scoring paragraphs from the validation sample (after pre-processing).
\begin{center}
    [Insert Table \ref{tab:doc2vec_parameters} here.]
\end{center}

\subsubsection{Cosine similarity}
\label{ch:cosine_similarity}

We use cosine similarity to measure the distance between the vector representations of two paragraphs, \textit{i.e.}, the cosine of the angle between the two vectors. As explained in \cite{AdosoglouLombardoPardalos2021}, cosine similarity is the most effective similarity estimate as the orientation of the embedding vectors is more stable than their magnitude due to the random initialization of the weights of the neural networks. Cosine similarity is a number between $-1$ and $1$. The closer the vectors, the higher the value. The similarity between two paragraphs, with vector representations $v_1$ and $v_2$, is therefore computed as $sim = \frac{v_1 \cdot v_2}{\| v_1 \| \hspace{0.2mm} \| v_2 \|}$.


\subsubsection{Cyber risk score}
\label{ch:cyber_score}

The cyber risk score is based on the cosine similarities with the cybersecurity descriptions from MITRE ATT\&CK. First, we compute the vector representation of every paragraph of every 10-K statement using the trained doc2vec best-performing specification. We also compute the vector representation of every sub-technique description from MITRE ATT\&CK. Next, we compute the cosine similarity of each paragraph from the 10-K statements with each of the MITRE descriptions. This gives 785 similarities for each paragraph from the 10-K statements. The cyber risk score of a paragraph is the maximum value out of those 785 similarities. Finally, we compute the score of a 10-K statement as the average score of the 1\% of its highest-scoring paragraphs.

This algorithm assumes that a typical 10-K statement has at most six or seven cyber risk-related paragraphs, representing 1\% of the paragraphs (average of 638 paragraphs per 10-K statement). This method has several advantages. First, taking a percentage of the total number of paragraphs, as opposed to a fixed number of paragraphs, makes it possible to have a meaningful comparison between 10-K statements that are much shorter or much longer than the others. Second, considering only the highest-scoring paragraphs makes the cyber risk score of the 10-K statement solely dependent on paragraphs that are most likely to be cyber risk-related.

Since the cosine similarity takes values between $-1$ and $1$, we only consider positive values for several reasons. First, interpreting a paragraph whose context is \quotes{opposite} to cybersecurity is not intuitive. Upon inspection of the paragraphs with negative values in the validation sample, we cannot uncover meaningful differences between paragraphs with negative scores and those with scores close to zero. Furthermore, only considering positive similarities guarantees that the cyber risk scores are between 0 and 1, making them comparable to those obtained using dictionary methods such as in \cite{JamilovReyTahoun2023W,FlorackisLoucaMichaelyWeber2023}.\footnote{We report in Appendix Table \ref{tab:negative_paragraphs} an excerpt of these negative paragraphs to illustrate that they are orthogonal to the concept of cyber risk and not indicators of cybersecurity.}

\begin{figure} 
    \noindent\makebox[\textwidth]{%
    \includegraphics[width=\textwidth]{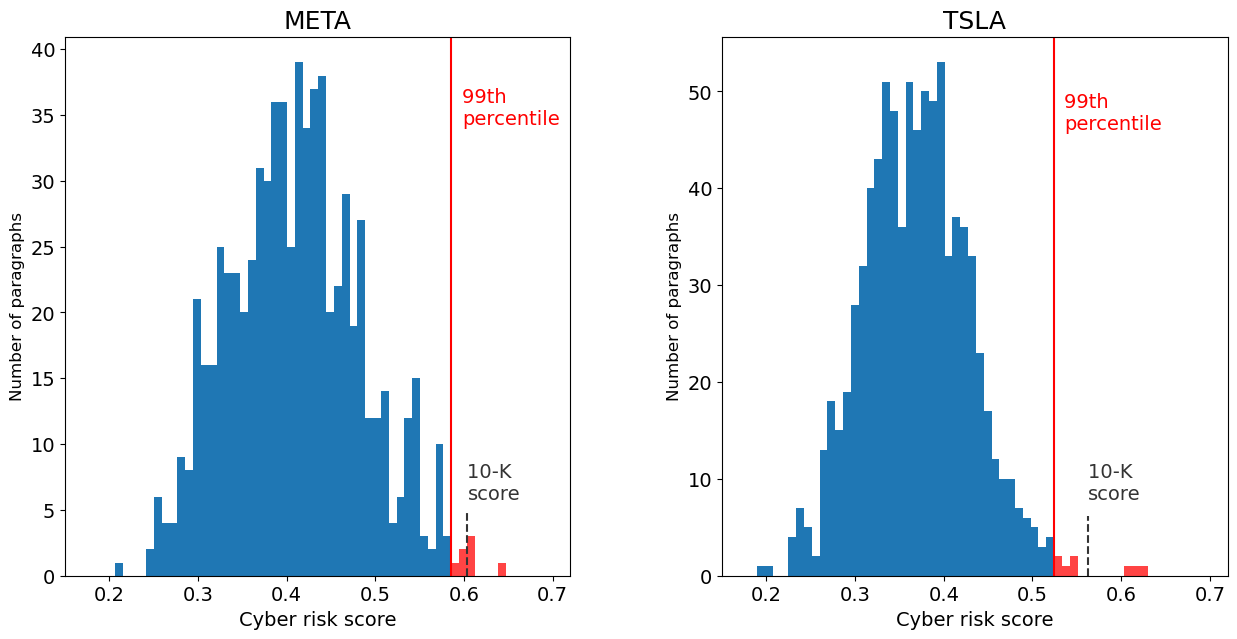}
    }
     \caption{ \textbf{Paragraph level score distributions for Meta Platforms and Tesla}}\bigskip
     \footnotesize{This figure depicts the distribution of paragraphs from 10-K statements filed in 2022 for Meta (left panel) and Tesla (right panel) based on their cyber risk score. We choose 50 bins and color in red the paragraphs within the top 1\% of cyber risk scores.}
     \label{fig:paragraph_score_distribution}
\end{figure}

Figure \ref{fig:paragraph_score_distribution} shows the distribution of the cyber risk scores of the paragraphs from the 10-K statements of Meta Platforms, Inc. and Tesla, Inc. filed in 2022. The paragraphs in red are the top 1\% of paragraphs with the highest cyber risk scores. We compute the 10-K cyber risk scores as the average score of this highest percentile.\footnote{We try, in turn, other scores, such as the firm average of the cyber risk score distribution with virtually no changes. We retain the top percentile approach since, intuitively, these paragraphs are the most likely to discuss actual cybersecurity issues.} This yields a score of $0.605$ for META and $0.563$ for TSLA.
\begin{center}
    [Insert Figure \ref{fig:paragraph_score_distribution} here.]
\end{center}

\subsection{Asset pricing tests}
\label{section:asset_pricing_tests}

We use two-step Fama-MacBeth regressions(\citealp{FamaMacBeth1973}) as follows. First, we estimate security betas on several factors using time series regressions with 2-year rolling windows. Second, we sort firms into 20 value-weighted portfolios based on their cyber risk. We compute the factor exposures of the portfolios and standardize the portfolio betas for economic interpretation. Finally, we estimate gammas using cross-sectional regressions of the portfolio returns on their lagged factor exposures and their value-weighted cyber risk score.

Next, we use the time series approach of \cite{GibbonsRossShanken1989} (hereafter, GRS) to test for portfolio efficiency. The GRS statistics allow testing whether the pricing errors are jointly equal to zero when using a model with several traded factors. We also use 20 portfolios built on cyber risk scores and factor betas, in turn, as test assets. We implement the GRS test and compare two model specifications, the five-factor model from \cite{FamaFrench2015} and the same five factors plus the cyber risk factor.

We implement the Bayesian approach of \cite{BarillasShanken2018}. Using this method, it is possible to compute the probability that a given factor model is best to price factor returns. The method does not presume that any model under consideration exactly satisfies the requirement that all alphas are zero, as it is possible that some relevant factors still need to be identified. The approach compares the relative success of the models in predicting the observed data. The method is based on \cite{BarillasShanken2017} and looks at the extent to which a model prices the factors left out and not the extent to which the model prices test assets. The unrestricted factor model is,
\begin{equation}
    R_t = \alpha +\beta F_t +\epsilon_t, \hspace{2mm}\epsilon_t \sim N(0,\Sigma),
\end{equation}
and the null hypothesis is $H_0: \alpha = 0$. The prior for $\alpha$ is concentrated at zero under the null hypothesis. Under the alternative, they assume a multivariate normal informative prior for $\alpha$: $P(\alpha \vert \beta, \Sigma) = MVN(0,k\Sigma)$, where $k$ reflects the beliefs about the potential magnitude of deviations from the expected return relation. By assumption, all the models contain the market factor. The marginal likelihood of a model is given by,
\begin{equation}
    ML = ML_U(F \vert Mkt) \times ML_R(F^* \vert Mkt, F) \times ML_R(R \vert Mkt, F, F^*),
    \label{eq:marginal_likelihood}
\end{equation}
where $ML_U$ is the unrestricted regression marginal likelihood, $ML_R$ is the restricted regression marginal likelihood ($\alpha$ constrained to zero), $F$ are the included factors and $F^*$ are the excluded factors. The $ML_U(X \vert Y)$ notation assumes the following regression equation:\\
$X_t = \alpha +\beta Y_t +\epsilon_t$. The unrestricted and restricted regression marginal likelihoods are given by,
\begin{equation}
\begin{split}
    ML_U &= \vert Y'Y \vert ^{-N/2} \vert S \vert ^{-(T-K)/2}Q \\
    ML_R &= \vert Y'Y \vert ^{-N/2} \vert S_R \vert ^{-(T-K)/2},
\end{split}
\end{equation}
where $\vert S \vert$ and $\vert S_R \vert$ are the determinants of the $N \times N$ cross-product matrices of the OLS residuals, $T$ is the number of periods, $K$ the number of factors, and $N$ the number of portfolios. The scalar $Q$ is given by,
\begin{equation}
    Q = \left( 1+ \frac{a}{a+k}(W/T) \right)^{-(T-K)/2} \left(1+\frac{k}{a} \right)^{-N/2},
\end{equation}
where $a = \nicefrac{(1+Sh(Y)^2)}{T}$, $k = \nicefrac{(Sh_{max}^2-Sh(Y)^2)}{N}$, $W$ is the GRS F-statistic times $\nicefrac{NT}{(T-N-K)}$, $Sh(Y)^2$ the squared sample Sharpe Ratio. Under the alternative prior, $k$ is the expected increment to the squared Sharpe ratio from the addition of one more factor. $Sh_{max}$ is the maximum expected Sharpe ratio. \cite{BarillasShanken2018} take $Sh_{max} = 1.5 \times Sh_{Mkt}$, which corresponds to a square root of the prior expected squared Sharpe ratio for the all factors-tangency portfolio 50\% higher than the market Sharpe ratio. They call this value the prior multiple. Similarly, we use 1.5 as the baseline value for the prior multiple but experiment with several values as in the original paper. The posterior probabilities, conditional on the data $D$, are given by Bayes' rule,
\begin{equation}
   P(M_j \vert D) = \frac{ML_j \times P(M_j)}{ \sum\limits_i ML_i \times P(M_i)},
\end{equation}
where $P(M_j)$ is the prior probability of the model. \cite{BarillasShanken2018} use uniform prior probabilities to avoid favoring one model over another. Hence, they cancel out in the division and can be omitted. $ML_R(R \vert Mkt, F, F^*)$, in equation \ref{eq:marginal_likelihood}, is the same for all combinations of $F$ and $F^*$. Hence, it also cancels out in the division and can be omitted.

Following the methodology of \cite{BarillasShanken2018}, we compute the posterior probabilities for each month from January 2010 until December 2022. We use all of the data available from January 2009 until the given time for each computation.

\section{Results}
\subsection{Cyber risk score}

Table \ref{tab:descriptive_statistics} presents descriptive statistics of the cyber risk score and various firm characteristics. The average cyber risk is 0.52, and its distribution is positively skewed, meaning there are more very high-risk firms than very low-risk ones. The cyber risk distribution is narrow, with a standard deviation of 0.03 and a spread between the top and bottom percentiles of 0.14. The correlation coefficients between our cyber risk score and firms' characteristics are small, except for Tobin's Q (0.23) and the Firm age (-0.17). The latter coefficient is consistent with the view that older firms are less subject to cyberattacks since their core businesses are less likely to be IT-related. Given that all other coefficients are below 0.15 in absolute value, we are confident that our score is orthogonal to other characteristics known to price stock returns.
\begin{center}
    [Insert Table \ref{tab:descriptive_statistics} here.]
\end{center}

\subsubsection{Time series and industry properties}

\begin{figure}
    \noindent\makebox[\textwidth]{%
    \includegraphics[width=\textwidth]{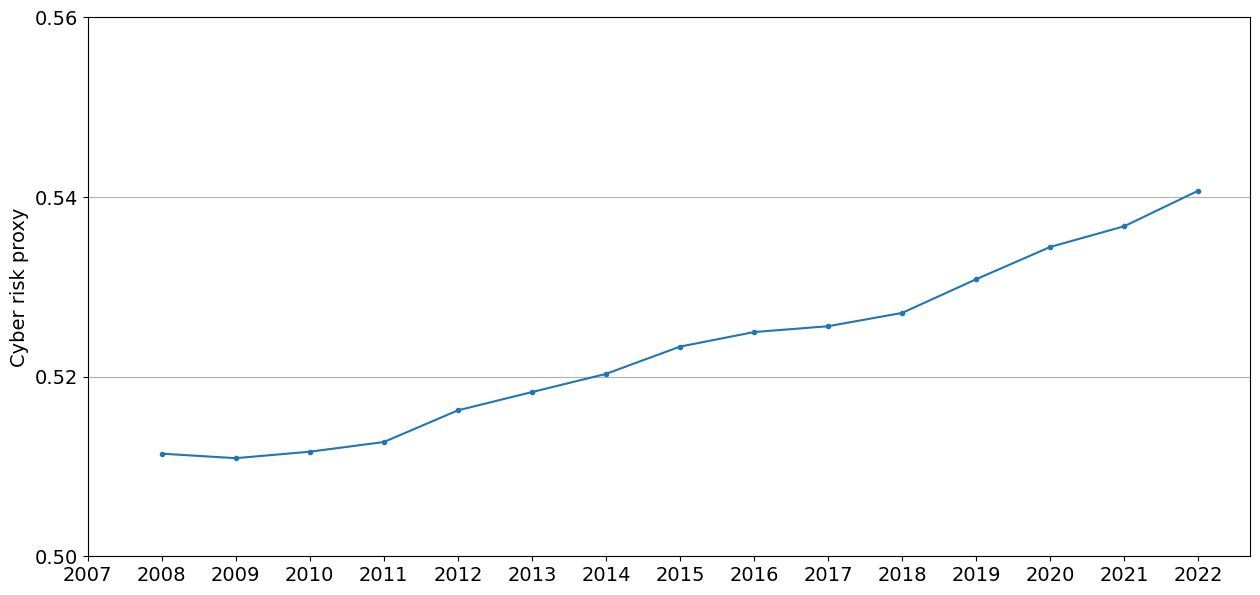}}   
    \caption{\textbf{Evolution of the average cyber risk across all firms}}\bigskip
    \footnotesize{This figure depicts the yearly evolution of our firm-level cyber risk measure averaged across all firms over our sample period: 2007--2022.}
    \label{fig:average_cyber_risk}
\end{figure}

\begin{figure}
    \noindent\makebox[\textwidth]{%
    \includegraphics[width=\textwidth]{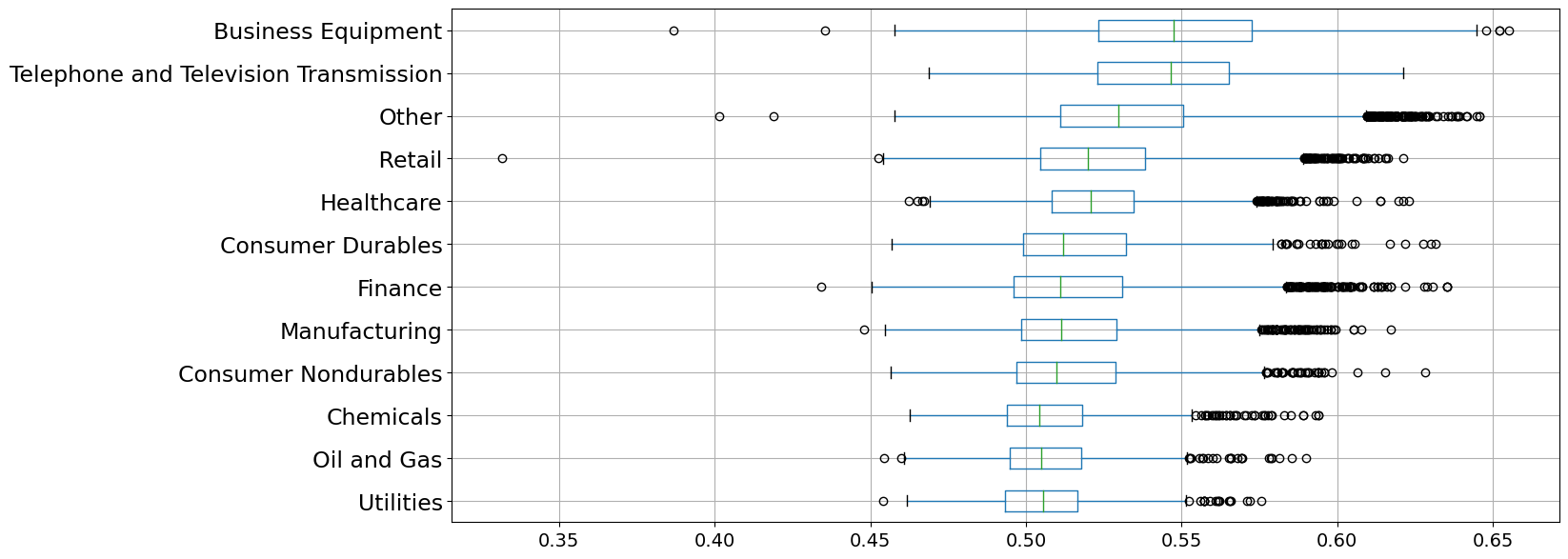}}
    
    \caption{\textbf{Cyber risk across industries}}\bigskip
    \footnotesize{This box plot depicts the minimum, first quartile, median, third quartile, and maximum values of our cyber risk measure for each Fama-French 12 industry classification. The whiskers extend no more than 1.5 times the interquartile range (third quantile - first quantile) from the edges of the box, ending at the farthest data point within that interval. Outliers are plotted as separate dots. Standard Industrial Classification (SIC) codes are obtained from CRSP. The conversion table, from SIC codes to the 12 Fama-French industries, is available on the \href{https://mba.tuck.dartmouth.edu/pages/faculty/ken.french/Data_Library/det_12_ind_port.html}{Kenneth French data repository}.}
    \label{fig:cyber_score_industries}
\end{figure}

Figure \ref{fig:average_cyber_risk} presents the cross-sectional average cyber risk score for every year in the study sample. We observe a monotonic positive time trend in line with the results of \cite{FlorackisLoucaMichaelyWeber2023} and \cite{JamilovReyTahoun2023W}.
\begin{center}
    [Insert Figure \ref{fig:average_cyber_risk} here.]
\end{center}
Figure \ref{fig:cyber_score_industries} shows the average cyber risk by industry, using the Fama-French 12 industry classification (also see, \citealp{FamaFrench1997}). Industries that rely on technology systems such as \quotes{Business Equipment} and \quotes{Telephone and Television Transmission} have high cyber risk. In contrast, sectors such as \quotes{Oil and Gas} and \quotes{Chemicals}, which traditionally rely less on technology systems, have lower scores. Nonetheless, the variation of cyber risk across industries remains limited, similar to the overall cyber risk distribution.
\begin{center}
    [Insert Figure \ref{fig:cyber_score_industries} here.]
\end{center}

\subsubsection{Determinants of firm-level cyber risk}

To investigate the dependence of cyber risk on firm characteristics, we perform two regressions, presented in Table \ref{tab:determinants_cyber_security}. In Model 1, we control for year- and firm-fixed effects; in Model 2, we control for year- and industry-fixed effects. In both models, firm age has a statistically significant negative coefficient at the 1\% level, reinforcing the view that younger firms are exposed to a higher cyber risk. The book-to-market coefficient is negative and significant at the 1\% level, meaning value firms have a lower cyber risk than growth firms. In Model 2, the coefficient of intangible assets to total assets is positive and statistically significant at the 1\% level. This supports the view that firms with more intangible assets, such as patents or software, have a higher cyber risk. The R-squared is low for both models, showing that firm characteristics can not readily explain cyber risk.
\begin{center}
    [Insert Table \ref{tab:determinants_cyber_security} here.]
\end{center}

\subsection{Univariate portfolio sorts}
\label{ch:univariate_sorts} 

\begin{figure}
    \noindent\makebox[\textwidth]{%
    \includegraphics[width=\textwidth]{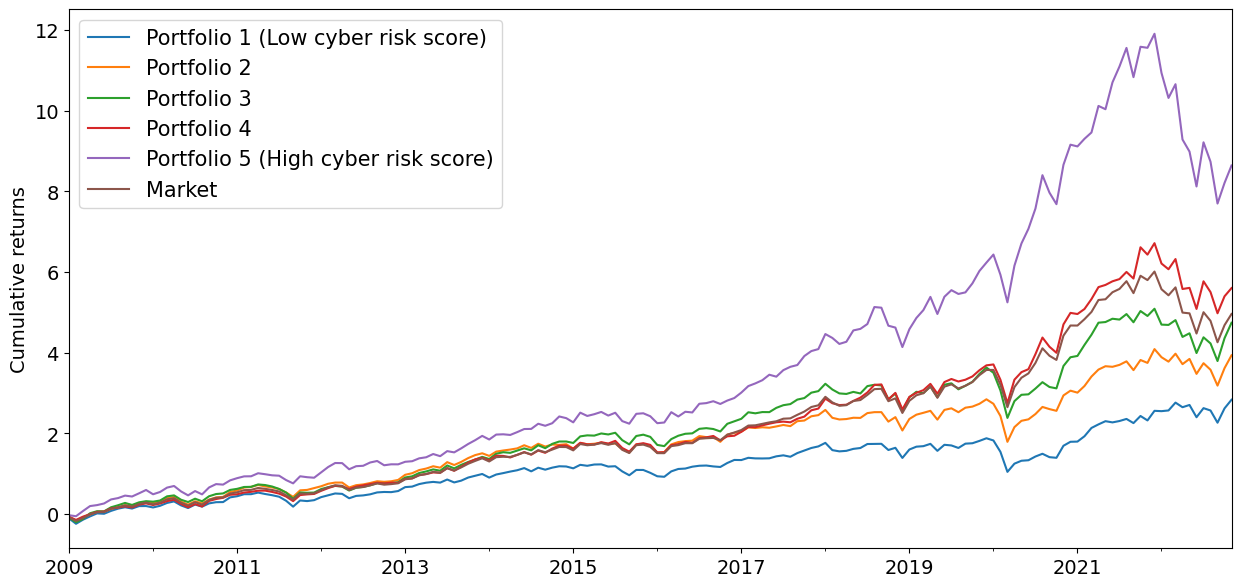}}
     \caption{\textbf{Cyber risk-sorted portfolios cumulative returns}}\bigskip
     \footnotesize{This figure depicts the performance of value-weighted portfolios of firms sorted on our cyber risk measure. We rebalance the portfolios quarterly. \quotes{Market} refers to the market portfolio obtained from the Kenneth French data repository. One unit of the ordinate axis unit is 100\% return. The frequency is monthly and the period is 2009--2022.}
     \label{fig:portfolio_cumulative_returns_full_sample}
\end{figure}

We sort firms into portfolios based on their cyber risk and study the returns on the portfolios. We use a firm characteristic approach based on the results of \cite{DanielTitman1997}, who argue that characteristics and not covariances determine expected returns. More precisely, we assign firms to five portfolios based on the cyber risk score of their most recent 10-K statement. We rebalance the portfolios quarterly to allow for listings and delistings and incorporate information from new 10-K statements. We build five value-weighted portfolios, where Portfolio 1 (5) is the low (high) cyber risk portfolio. We track the performance of the portfolios from January 2009 until December 2022. Figure \ref{fig:portfolio_cumulative_returns_full_sample} shows the evolution of the cumulative returns of the market and the five portfolios. We observe that the higher the cyber risk of the portfolio, the higher the cumulative returns. Specifically, Portfolio 5 significantly outperforms the market.
\begin{center}
    [Insert Figure \ref{fig:portfolio_cumulative_returns_full_sample} here.]
\end{center}
Table \ref{tab:excess_returns_alphas_full_sample} presents the excess returns and alphas of the portfolios with respect to three standard factor models. The average monthly excess returns increase monotonically from 0.88\% to 1.44\%, from low to high cyber risk portfolios. The long-short portfolio, going long in Portfolio 5 and short in Portfolio 1, has excess returns and alphas that remain statistically significant at the 5\% level, after controlling for the \cite{FamaFrench2015} five-factor model. In addition, we also observe that the Sharpe, Treynor, and Sortino ratios all experience monotonical growth as the cyber risk increases, reinforcing our views that the performance is not driven by compensation for total volatility, market exposure, or downside risk, respectively.\footnote{in Appendix, Figure \ref{fig:rolling_correlation} we depict the 2-year rolling window correlation of the cyber risk long-short portfolio and the market. On average, this correlation shows up at only 2\%.}
\begin{center}
    [Insert Table \ref{tab:excess_returns_alphas_full_sample} here.]
\end{center}

\subsection{Double sorts}

We also perform double sorts, where we first sort five portfolios on an alternative firm characteristic and then sort the resulting portfolios on cyber risk. Akin to the previous section, we build value-weighted portfolios that we rebalance quarterly. We use three characteristics for the first sort, market beta, book-to-market ratio, and firm size. Table \ref{tab:double_sorts} shows the average excess returns of each portfolio. The factor structure of cyber risk is more prevalent among large firms, low to medium book-to-market firms, and medium beta firms. The long-short portfolio has positive excess returns for most of the portfolio sorts. This further supports the view that cyber risk captures a specific variation in average returns by controlling for market beta, firm size, and book-to-market value.
\begin{center}
    [Insert Table \ref{tab:double_sorts} here.]
\end{center}

\subsection{Asset pricing tests}

\subsubsection{Fama-Macbeth regressions}

Table \ref{tab:Fama_MacBeth} presents the results of Fama-Macbeth regressions. Specification 1 only includes the market factor. The coefficient on the market is insignificant, and the average adjusted R-squared is small, showing that the CAPM can not price the cyber beta-sorted portfolios. In Specification 2, we use the cyber risk proxy, and as shown, the risk premium is statistically significant, and the average adjusted R-squared increases from virtually 0 to 0.13. Specifications 3, 4, and 5 control for other common factors, while the cyber risk premium remains economically and statistically significant. The economic interpretation of this table is that a one-standard-deviation increase in cyber risk drives returns up by about 18 basis points per month. This increase is statistically significant at the 10 or 5\% level, even when controlling for other common factors.
\begin{center}
    [Insert Table \ref{tab:Fama_MacBeth} here.]
\end{center}
We further test whether an industry effect does not drive the cyber risk premium. We run Fama-Macbeth regressions, expanding the set of explanatory variables with betas on seven value-weighted Fama-French industry-portfolios\footnote{We drop four of them due to collinearity and leave the \quotes{Other} categories aside because of the lack of interpretability.}. We report these results in Specification 6 of Table \ref{tab:Fama_MacBeth}. The cyber risk premium decreases to 0.150\% but remains statistically significant at the 10\% level. These results are obtained using portfolios sorted on cyber risk. As highlighted by \cite{LewellenNagelShanken2010}, asset pricing tests, such as Fama-Macbeth regressions, can be improved by using additional portfolios sorted on other firm characteristics. Following their recommendations, we expand the set of test assets with portfolios sorted by industry. We report the results in Specification 7 of Table \ref{tab:Fama_MacBeth}. Once again, the economic significance of the cyber risk premium is virtually unchanged at 0.176\% per month, and its statistical significance remains below the 5\% level.

\subsubsection{GRS test}
\label{ch:GRS_results} 

We implement the GRS test as follows. We build 20 value-weighted portfolios sorted on cyber risk. Next, we compute the GRS test statistic using the five-factor model from \cite{FamaFrench2015} and the same model augmented with the cyber risk factor. We report the results and repeat this procedure, sorting portfolios on market beta, firm size, and book-to-market ratio. Table \ref{tab:GRS} presents the results. The GRS test statistic is smaller for the model containing the cyber risk factor when sorting on cyber risk, size, and book-to-market. Interestingly, when sorting on firm size and book-to-market, we reject the null hypothesis that $\alpha_i = 0, \hspace{2mm} \forall i$, for the five-factor model but not for the model containing the cyber risk factor. This is not the case when sorting on market beta. However, we can not reject the null hypothesis for either model.
\begin{center}
    [Insert Table \ref{tab:GRS} here.]
\end{center}

\subsubsection{Bayesian factor model selection}

\begin{figure}
    \noindent\makebox[\textwidth]{%
    \includegraphics[width=\textwidth]{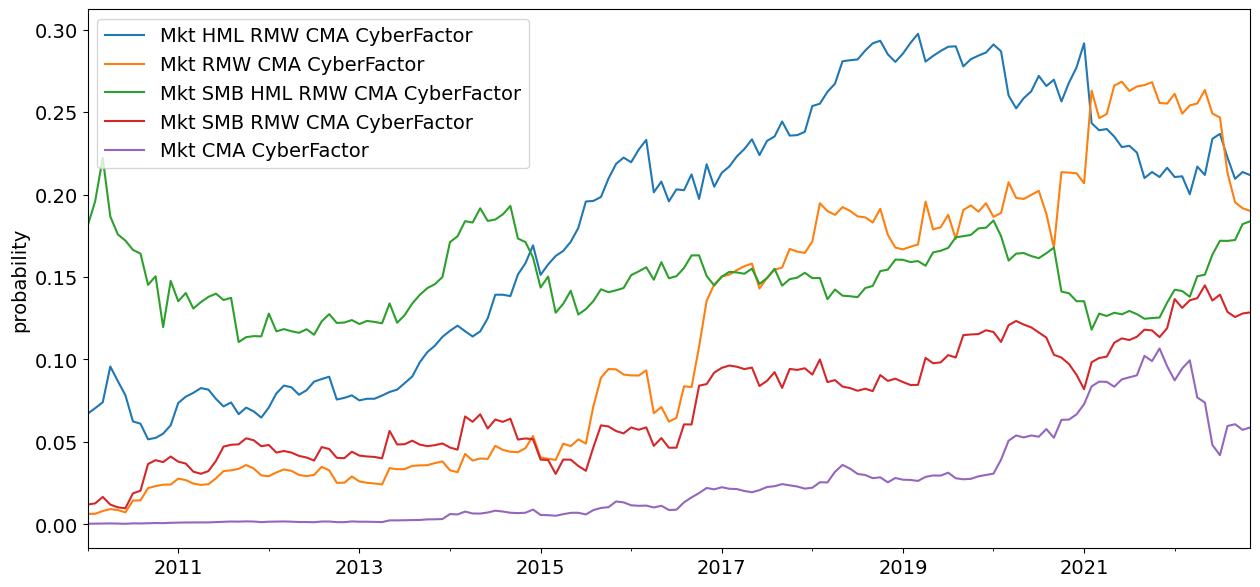}}
     \caption{\textbf{Factor model posterior probabilities}}\bigskip
     \footnotesize{The figure depicts the posterior probabilities for the top five models, ranked at the end of the sample. \quotes{Mkt} refers to the excess return on the market from the Kenneth French data repository. HML and SMB refer to the book-to-market and size factors of \cite{FamaFrench1992}. CMA and RMW refer to the investment and operating profitability factors of \cite{FamaFrench2015}. \quotes{CyberFactor} refers to the long-short portfolio built on our cyber risk score. The prior multiple is 1.5, and the study period is from January 2010 to December 2022.}
     \label{fig:factor_model_probabilities}
\end{figure}

\begin{figure}
    \noindent\makebox[\textwidth]{%
    \includegraphics[width=\textwidth]{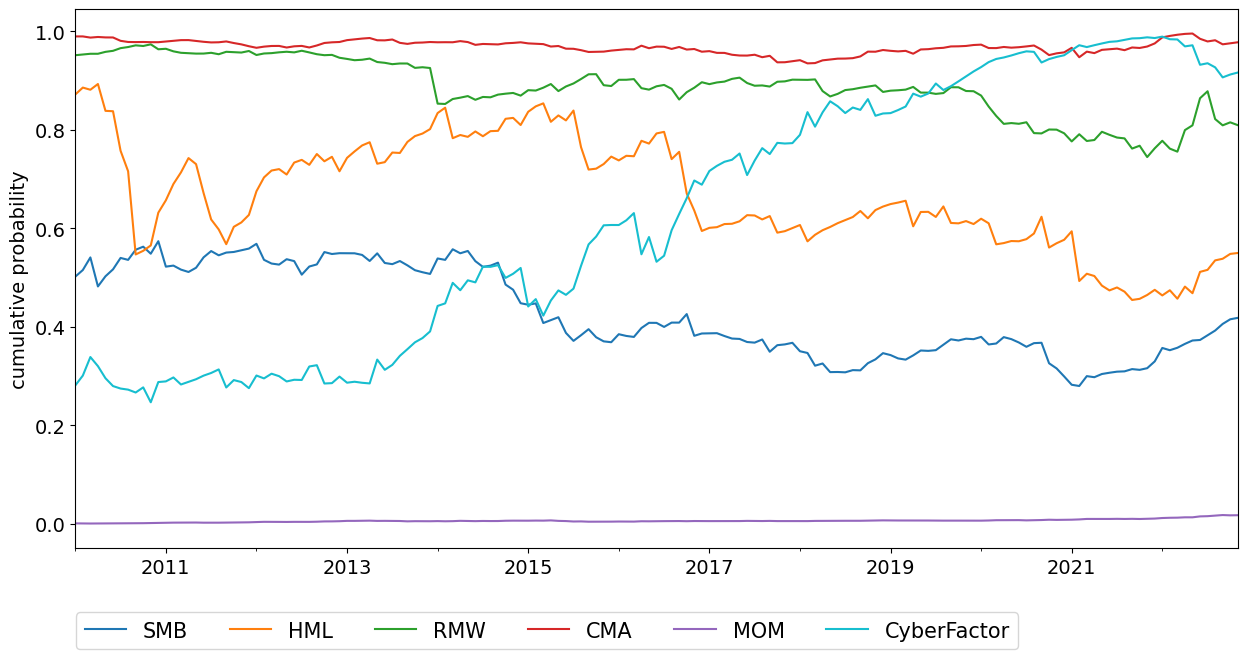}}
     \caption{\textbf{Cumulative posterior factor probabilities}}\bigskip
     \footnotesize{This figure depicts the cumulative posterior probabilities, \textit{i.e.} the sum of probabilities of all models containing the factor. HML and SMB refer to the book-to-market and size factors from \cite{FamaFrench1992}. MOM refers to the momentum factor from \cite{Carhart1997}. CMA and RMW refer to the investment and operating profitability factors from \cite{FamaFrench2015}. \quotes{CyberFactor} refers to the long-short portfolio built on our cyber risk score. The prior multiple is 1.5, and the study period is from January 2010 to December 2022.}
     \label{fig:cumulative_factor_probabilities}
\end{figure}

Given the results of the GRS tests, a subset of factors could explain the cross-section of returns. The approach of \cite{BarillasShanken2018} detailed in Section \ref{section:asset_pricing_tests} allows us to determine the optimal combination of factors. Including the market factor in all cases, we compute the 62 permutations between all factors, \textit{i.e.} HML, SMB, CMA, RMW, MOM, and the cyber factor.\footnote{There are $2^6 - 2$ permutations since we discard the case when all six factors are included in the model.} Figure \ref{fig:factor_model_probabilities} presents the posterior probabilities of the five most likely models ranked at the end of the sample. All five optimal models contain the cyber risk factor, and the model with the highest probability, of 21.18\%, comprises the market, book-to-market, investment, operating profitability, and cyber risk factors.
\begin{center}
    [Insert Figure \ref{fig:factor_model_probabilities} here.]
\end{center}
Figure \ref{fig:cumulative_factor_probabilities} presents the cumulative factor probabilities, \textit{i.e.} the sum of probabilities of all models containing the factor. The cyber risk factor has a cumulative probability of 91.66\% at the sample's end. Unlike the remaining factors, the investment and operating profitability factors also have high cumulative probabilities.
\begin{center}
    [Insert Figure \ref{fig:cumulative_factor_probabilities} here.]
\end{center}
Finally, we study the sensitivity of the model probabilities to the prior multiple of the Sharpe ratio. Following \cite{BarillasShanken2018}, we repeat the analysis using three other values of prior multiple: 1.25, 2, and 3. Table \ref{tab:prior_sensitivity} reports the posterior model probabilities at the end of the sample for the top five models for each prior. The set of top five models is unchanged for each prior multiple. Furthermore, the book-to-market factor is no longer included in the most likely model for higher values of the prior multiple.
\begin{center}
    [Insert Table \ref{tab:prior_sensitivity} here.]
\end{center}
\subsection{Comparison with the work of \cite{FlorackisLoucaMichaelyWeber2023}}

To compare our cyber risk score to the one of \cite{FlorackisLoucaMichaelyWeber2023}, we start by computing the correlation between the two and obtain 0.34.\footnote{The \cite{FlorackisLoucaMichaelyWeber2023} data is available at:\\ \href{https://tinyurl.com/4bcr68xf}{https://alucutac-my.sharepoint.com/personal/christodoulos\_louca}.} Encouragingly, the correlation is positive but relatively low, which shows that our estimate is novel.

The methodology of this paper improves that of \cite{FlorackisLoucaMichaelyWeber2023}. Specifically, the Paragraph Vector algorithm is more potent than a dictionary approach in capturing cyber risk-related content in 10-K statements. Indeed, the dictionary will only contain some words needed to identify all cyber risk-related text. The dictionary used by \cite{FlorackisLoucaMichaelyWeber2023} can locate many of these texts unless the authors of the 10-K statements use slightly different wording than the others, which is not uncommon. For example, Appendix Table \ref{tab:missed_paragraphs} shows three paragraphs that were missed by their approach, and consequently, the firm-year observations got a cyber risk score of zero. Using the methodology presented in this paper, these 10-K statements were given an above-average cyber risk score.

To investigate this further, we use our cyber risk score to sort firms attributed with a zero cyber risk score by \cite{FlorackisLoucaMichaelyWeber2023}) into three portfolios. We report the portfolios' average excess returns and alphas in Table \ref{tab:zero_risk_firm_sorts}. We observe that there is a structure left to exploit in returns sorted from zero-risk firms with dictionary methods. The average excess returns and alphas increase with the cyber risk (as measured using our indicator), and the long-short portfolio has economically significant excess returns and alphas.
\begin{center}
    [Insert Table \ref{tab:zero_risk_firm_sorts} here.]
\end{center}
An additional advantage of systematically capturing a score is that it resolves the unbalanced portfolio's firm number that their method was implying, thereby allowing the formation of as many portfolios as needed.
Finally, we test how much cyber risk information is missed in 10-K statement parts other than Items 1.A. We sort portfolios based on the cyber score that excludes this section, and we report the results in Table \ref{tab:excess_returns_alphas_no_1A}. We observe that alphas and statistical significance levels are slightly lower, but our interpretation is unchanged. Thus, we conclude that the entirety of the filings needs to be considered since a significant amount of cyber risk-related information remains outside of Items 1.A. Specifically, computing other statistics, we find that, on average, only 34\% of paragraphs used for our cyber risk score comes from Items 1.A. This proportion has increased regularly over the years, from about 15\% in 2008 to about 50\%  in 2022. We also report an excerpt of these Paragraph Vector-identified top-scoring paragraphs in Appendix Table \ref{tab:missed_paragraphs}.
\begin{center}
    [Insert Table \ref{tab:excess_returns_alphas_no_1A} here.]
\end{center}
In unreported tests, we also circumvent the cyber risk score to be only on Item 1.A; we obtain slightly less significant results, further supporting that the information should be extracted from the full statement.

\subsection{Robustness tests}

\subsubsection{Long-run cyber risk}

By design, our cyber risk score depends only on each company's most recent 10-K statement. A firm may discuss cybersecurity concerns and risks extensively in its 10-K statement in year $T$, for example, because of an increasing number of cyberattacks in the industry, resulting in a high cyber risk score. Having focused on cybersecurity in year $T$ and not having been attacked itself, the firm could decide not to talk about cyber risks in its following 10-K statements (in years $T+i$) or not as much as in year $T$, even though it still has similar cyber risks. The previously computed cyber risk score would miss these cases as it has no memory.

We compute the expanding average cyber risk score to study the long-run cyber risk of firms. The long-run cyber risk score in year $T$ is the average of its simple cyber risk scores from 2008 to year $T$. This new estimate could account for the case companies described above. The advantage of using the long-run average instead of the long-run maximum is that it does not discard the observations. This is beneficial when considering firms in the following situation: consider a firm that discusses cyber risks in its 10-K statement in year $T$ following a cyberattack or data breach. The firm might purchase protection (insurance or software), minimize its future cyber risk, and not discuss this risk in its subsequent 10-K statements. The low cyber risk scores in the upcoming years represent the firm's reality and should not be discarded.

We repeat the portfolio sorts from Section \ref{ch:univariate_sorts} using the long-run cyber risk to sort firms. Table \ref{tab:excess_return_alphas_long_run} presents the results. We observe no significant change from Table \ref{tab:excess_returns_alphas_full_sample}, and the long-short portfolio remains significant at the 5\% or 10\% level. These results indicate that firms incorporate all available information regarding their cyber risk in their newest 10-K statements; hence, incorporating information from past statements does not improve the estimation of the cyber risk.
\begin{center}
    [Insert Table \ref{tab:excess_return_alphas_long_run} here.]
\end{center}

\subsubsection{Controlling for cybersecurity firms}
\label{subsection:controlling_for_cybersecurity_firms}

Our cyber risk score does not make a distinction between firms that discuss cybersecurity because they consider it a risk and firms that are cybersecurity solutions providers, for example, Fortinet\footnote{Available at: \href{https://www.fortinet.com/}{https://www.fortinet.com/}}. As there is no dedicated cybersecurity industry classification, we identify cybersecurity firms using the HACK ETF\footnote{Available at: \href{https://etfmg.com/funds/hack/}{https://etfmg.com/funds/hack/}}. The fund's description explains that this ETF invests in companies providing cybersecurity solutions, including hardware, software, and services. As cybersecurity providers, these firms are expected to discuss cybersecurity extensively in their 10-K statements, resulting in a false high cyber risk score. Indeed, these firms have an average score of 0.59, which is in the top 3\% of cyber risk scores. We repeat the analysis from Section \ref{ch:univariate_sorts}, and we exclude the holdings of the HACK ETF from the universe of firms.

Table \ref{tab:excess_return_alphas_no_cyber} presents the results. The results for Portfolios 1, 2, and 3 are unchanged, and the excess returns and alphas of Portfolios 4 and 5 increase. Furthermore, we observe that the t-statistics on Portfolios 4 and 5 also increase.
\begin{center}
    [Insert Table \ref{tab:excess_return_alphas_no_cyber} here.]
\end{center}

\subsubsection{Before and after \cite{FlorackisLoucaMichaelyWeber2023} was first released}

We repeat our analysis by splitting the study period before and after the time of the first release of \cite{FlorackisLoucaMichaelyWeber2023} on SSRN (January 2009 until October 2020) and then after the release (November 2020 until December 2022). Table \ref{tab:excess_return_alphas_before} presents the results using the period before the publication. The long-short portfolio has statistically significant positive excess returns and alphas (significant at the 1\% level). The outperformance of Portfolio 5 and the underperformance of Portfolio 1 is also more substantial, with the long-short portfolio having an average monthly excess return of 0.94\%. Portfolio 1 has statistically significant negative alphas at the 1\% level.
\begin{center}
    [Insert Table \ref{tab:excess_return_alphas_before} here.]
\end{center}
Table \ref{tab:excess_return_alphas_after} presents the results using the period after the release. Portfolio 1 has a high average monthly excess return of 2.16\%, significant at the 5\% level, and outperforms the other portfolios whose excess returns are not statistically significant. The long-short portfolio has negative average excess returns of -1.49\%, significant at the 5\% level, and statistically significant negative alphas when controlling for the market at the 1\% level. Still, the alphas are not statistically significant when controlling for the factors from \cite{Carhart1997} or \cite{FamaFrench2015}. Last, the correlation between the long-short cyber risk and the market portfolios spikes in this period (see Appendix, Figure \ref{fig:rolling_correlation}).
\begin{center}
    [Insert Table \ref{tab:excess_return_alphas_after} here.]
\end{center}
There could be several explanations for these results. It could be that the publication made some arbitrageurs trade stocks based on their cyber risk evaluation, which results in lower returns post-publication, as explained in \cite{McLeanPontiff2016}. Another interpretation provided by \cite{FlorackisLoucaMichaelyWeber2023} is that high-risk firms lost value in a higher concentration of cyber incidents while low-risk firms, being less affected, appreciated.

For instance, T-Mobile was a victim of a cyberattack in August 2021 during which more than 76.6 million current and former customers' information had been accessed\footnote{Available at:\href{https://www.t-mobile.com/news/network/cyberattack-against-tmobile-and-our-customers}{https://www.t-mobile.com/news/network/cyberattack-against-tmobile-and-our-customers}}. Furthermore, the U.S Treasury Department published a report that as of June 2021, financial institutions had already reported 635 suspicious ransomware-related activities which constituted a 30\% increase from all reported activity in 2020\footnote{Available at: \href{https://cyberscoop.com/ransomware-treasury-cryptocurrency-sanctions/}{https://cyberscoop.com/ransomware-treasury-cryptocurrency-sanctions/}}. The report also found that the cost of ransomware payments increased. These events could explain why Portfolio 5 has low returns and the long-short portfolio has negative returns. However, the post-publication period is much smaller; thus, the estimates could be spurious.

\section{Conclusion}
This paper implements the \quotes{Paragraph Vector} algorithm to estimate firms' cyber risk based on their 10-K statements. Documenting that this score is unrelated to other economic variables, we sort portfolios across this cyber risk score that we submit to various asset pricing tests. Our results support the view that cyber risk is priced in the cross-section of stock returns. Indeed, a long-short strategy on cyber risk sorted portfolios has a positive and statistically significant alpha over traditional factor models and an average monthly excess return of 0.56\%. We investigate the interactions of cyber risk with the other sources of risks by performing double sorts. The results show that cyber risk captures a variation in average returns by controlling for market beta, firm size, and book-to-market value. Furthermore, we show that cyber risk has a significant premium using Fama-Macbeth regressions. Using the GRS test of \cite{GibbonsRossShanken1989} and the methodology from \cite{BarillasShanken2018}, we show that the cyber risk factor helps to price stocks and is present in the five most likely factor models.

We compare our score to the one of \cite{FlorackisLoucaMichaelyWeber2023} and conclude that while being mildly correlated with theirs (34\%), our score performs better at capturing cyber risk. By design, our method allows us to capture cyber risk contexts and not only specific keywords. Consequently, this method can be unrestricted to specific sections of firms' disclosures, thereby integrating all available information. We confirm this with a specific test on firms sorted on 10-K statements, excluding Item 1.A and document excess returns in line with those obtained using the full corpus. Second, our approach always enables us to obtain a firm cyber risk score, avoiding the issue of classifying firms as zero-risk. Specifically, while sorting firms attributed to zero-cyber risk with the dictionary approach, we uncover a remaining structure in the portfolio's returns with tercile long-short portfolios' returns statistically significant at the 5\% level and of 0.53\% per month. A third benefit of this approach is that it does not require the use of past cyberattacks to be calibrated, which solves the problem of whether what is captured are \quotes{post-cyberattacks recovery}, or actual latent cyber risk.

Finally, we run two robustness tests. First, we compute the long-run cyber risk as the expanding average of the cyber risk. We perform the portfolio sorts and observe that the results are very similar. Hence, we conclude that firms incorporate all available information about cyber risk in their newest 10-K statement. Second, we exclude cybersecurity firms from the sample and perform the portfolio sorts. We find that the average monthly returns of the two high cyber-risk portfolios increase, and the others are left unchanged, supporting the view that the alpha arises from cyber risk exposure. We are confident that this new semantic approach incorporating information in textual documents would successfully identify other risk factors or characteristics, such as legislative or geopolitical risks. These risks could be readily captured by changing the MITRE knowledgebase to one close to the risk under scrutiny.



\clearpage
\bibliographystyle{styles/jfe}
\bibliography{bibliography.bib}

\clearpage
\section*{Figures and Tables}
\phantom{test}
\vspace{30mm}

\clearpage

\clearpage
\begin{figure}
    \centering
    \begin{tacticbox}
    Tactics represent the \quotes{why} of an ATT\&CK technique or sub-technique. It is the adversary's tactical goal: the reason for acting. For example, an adversary may want to achieve credential access, collect data, or run malicious code.
        \begin{idbox}
            Techniques represent \quotes{how} an adversary achieves a tactical goal by acting. For example, an adversary may dump credentials to achieve credential access or may use social engineering to run malicious code. 
            \begin{subidbox}
                Sub-techniques describe the different types of a technique. For example, when trying to run malicious code for social engineering, the adversary may use a malicious link, a malicious file, or a malicious image for execution.
            \end{subidbox}
        \end{idbox}
    \end{tacticbox}
    \caption{\textbf{Structure of the tactic descriptions on MITRE ATT\&CK}}
    \label{fig:MITRE_structure}
\end{figure}

\clearpage
\begin{table}
\noindent\makebox[\textwidth]{%
    \begin{tabular}{l|c|l}
    \hline
    \hline
    & & \multicolumn{1}{c}{Description}\\
    \hline
    Tactic & Credential Access & \multirow{3}{90mm}{Adversaries may forge web cookies that can be used to gain access to web applications or Internet services. Web applications and services (hosted in cloud SaaS environments or on-premise servers) often use session cookies to authenticate and authorize user access.}\\[3ex]
    Technique & Forge Web Credentials & \\[3ex]
    Sub-technique & Web Cookies & \\[3ex]
    \hline
    Tactic & Reconnaissance & \multirow{3}{90mm}{Adversaries may gather employee names that can be used during targeting. Employee names can be used to derive email addresses as well as to help guide other reconnaissance efforts and/or craft more-believable lures. }\\[2ex]
    Technique &  Gather Victim Identity Information & \\[2ex]
    Sub-technique & Employee Names & \\[2ex]
    \hline
    \hline
    \end{tabular}
}
\caption{\textbf{Examples of sub-technique descriptions from MITRE ATT\&CK}}
\label{tab:MITRE_examples}
\end{table}

\clearpage

\clearpage
\begin{table}
    \resizebox{\textwidth}{!}{
    \begin{tabular}{lcccccccc}
        & \textbf{Method} & \textbf{Training Size} & \textbf{Vector Size} & \textbf{Window Size} & \textbf{Min Count} & \textbf{Sub-Sampling} & \textbf{Negative Sampling} & \textbf{Epoch}\\
         \cmidrule(lr){2-2} \cmidrule(lr){3-3} \cmidrule(lr){4-4} \cmidrule(lr){5-5} \cmidrule(lr){6-6} \cmidrule(lr){7-7} \cmidrule(lr){8-8} \cmidrule(lr){9-9}
        Best specification & DBOW & 1.7M & 200 & 15 & 5 & $10^{-5}$ & 5 & 50 \\
        Baseline specification & DBOW  & - & 300 & 15 & 5 & $10^{-5}$ & 5 & 20 \\
    \end{tabular}}
    \caption{\textbf{doc2vec parameters}}\bigskip
    \footnotesize{This table reports the chosen parameters used to train the doc2vec specification and the baseline specification parameters of \cite{LauBaldwin2016}. The chosen parameters correspond to the best specification we find in computing the vector representations of the paragraphs from 10 randomly selected 10-K statements from 2008 (validation sample). We try each specification and compare the highest-scoring paragraphs between the specifications. We assume the best specification to be the one with the highest proportion of highest-scoring cyber risk-related paragraphs. \quotes{DBOW} stands for distributed bag-of-words.}
    \label{tab:doc2vec_parameters}
\end{table}

\clearpage

\clearpage
\begin{table}
\noindent\makebox[\textwidth]{%
    \begin{tabular}{lcccccccc}
    \hline
    \hline
     & Mean & SD & P1 & P25 & P50 & P75 & P99 & Correlation with\\
     & & & & & & & & cyber risk\\
    \hline
    Cyber risk & 0.52 & 0.03 & 0.47 & 0.50 & 0.52 & 0.54 & 0.61 & -\\
    Firm Size (ln) & 20.18 & 2.39 & 13.15 & 18.53 & 20.25 & 21.86 & 25.46 & -0.10\\
    Firm Age (ln) & 2.70 & 1.06 & -0.88 & 2.21 & 2.93 & 3.41 & 4.07 & -0.17\\
    ROA & -0.11 & 0.47 & -2.57 & -0.07 & 0.02 & 0.07 & 0.36 & -0.05\\
    Book to market ratio & 0.68  & 1.15 & 0.02 & 0.24 & 0.46 & 0.81 & 4.42 & -0.12\\
    Tobin's Q & 2.20 & 2.14 & 0.58 & 1.09 & 1.50 & 2.37 & 12.15 & 0.23\\
    Market Beta & 1.20 & 0.84 & -1.01 & 0.71 & 1.13 & 1.60 & 3.90 & 0.00\\
    Intangibles/Assets & 0.17 & 0.21 & 0.00 & 0.00 & 0.07 & 0.27 & 0.78 & 0.14\\ 
    Debt/Assets & 0.53 & 0.28 & 0.06 & 0.32 & 0.52 & 0.70 & 1.48 & -0.09\\
    ROE & -0.08 & 0.61 & -2.96 & -0.08 & 0.07 & 0.15 & 0.88 & -0.06\\
    Price/Earnings & 1.55 & 112.17 & -511.4 & -4.44 & 12.57 & 23.82 & 294.46 & -0.01\\
    Profit Margin & -0.38 & 5.53 & -25.20 & 0.21 & 0.36 & 0.57 & 0.94 & 0.00\\
    Asset Turnover & 0.92 & 0.74 & 0.01 & 0.39 & 0.76 & 1.26 & 3.54 & -0.03\\
    Cash Ratio & 1.85 & 3.41 & 0.01 & 0.23 & 0.65 & 1.81 & 18.20 & 0.11\\
    Sales/Invested Capital & 1.54 & 1.59 & 0.01 & 0.56 & 1.08 & 1.94 & 8.88 & -0.01\\
    Capitalization Ratio & 0.30 & 0.32 & 0.00& 0.02 & 0.24 & 0.47 & 1.54 & -0.10\\
    R\&D/Sales & 0.67 & 4.21 & 0.00 & 0.00 & 0.00 & 0.08 & 19.40 & 0.03\\
    ROCE & 0.00 & 0.45 & -1.97 & -0.02 & 0.09 & 0.17 & 0.95 & -0.07\\
    \hline
    \hline
    \end{tabular}}
    
\caption{\textbf{Descriptive statistics of the cyber risk measure and firm characteristics}}\bigskip
\footnotesize{This table reports firm-level characteristics, Mean, standard deviation, percentiles, and correlation with cyber-risk. The measure is winsorized at the 1\textsuperscript{st} and 99\textsuperscript{th} percentile (by year). We define the characteristics in Table \ref{tab:variable_descriptions}.}
\label{tab:descriptive_statistics}
\end{table}

\clearpage

\clearpage

\clearpage
\begin{table}
    \begin{adjustbox}{width=0.44\textwidth,center}
    \vspace{-6em}
    \begin{tabular}{lcc}
        \hline
        \hline
        \multicolumn{3}{c}{Firm-level indicator of cyber risk}\\
        \hline
        & Model 1 & Model 2\\
        \hline
        Constant & \textbf{-0.416}$^{***}$ & \textbf{-0.738}$^{***}$ \\
         & [-24.89] & [-14.22]\\
        Firm Size (ln) & \textbf{0.019} &  \textbf{0.024}\\
        & [0.56] & [1.22] \\
        Firm Age (ln)& \textbf{-0.114}$^{***}$ & \textbf{-0.211}$^{***}$\\
        & [-3.91] & [-12.09]\\
        ROA & \textbf{0.057} & \textbf{0.0321}$^{**}$\\
        & [0.79] & [2.27] \\
        Book to Market & \textbf{-0.023}$^{***}$ & \textbf{-0.066}$^{***}$\\
        & [-4.82] & [-3.67] \\
        Tobin's Q & \textbf{0.019}$^{***}$ & \textbf{0.112}$^{***}$\\
        & [2.84] & [7.87] \\
        Market Beta & \textbf{-0.009} & \textbf{-0.013}\\
        & [-1.54] & [-1.31] \\
        Intangibles/Assets & \textbf{-0.026}$^{**}$ & \textbf{0.082}$^{***}$\\
        & [-2.04] & [5.51] \\
        Debt/Assets & \textbf{-0.032}$^{**}$ & \textbf{0.032} \\
        & [-2.49] & [1.21]\\
        ROE & \textbf{0.002} & \textbf{-0.009} \\
        & [0.37] & [-0.72] \\
        Price/Earnings & \textbf{0.005} & \textbf{0.002} \\
        & [1.20] & [0.29] \\
        Profit Margin & \textbf{0.006} & \textbf{0.048}$^{***}$\\
        & [1.18] & [3.98] \\
        Asset Turnover & \textbf{-0.014} & \textbf{-0.135}$^{***}$\\
        & [-0.67] & [-4.49] \\
        Cash Ratio & \textbf{0.001} & \textbf{0.019} \\
        & [0.09] & [1.19] \\
        Sales/Invested Capital & \textbf{0.008} & \textbf{0.104}$^{***}$\\
        & [0.54] & [3.80]\\
        Capital Ratio & \textbf{0.001} & \textbf{-0.191}$^{***}$\\
        & [0.02] & [-8.49] \\
        R\&D/Sales & \textbf{-0.001} & \textbf{-0.003}\\
        & [-0.22] & [-0.30]\\
        ROCE & \textbf{0.005} & \textbf{0.000} \\
        & [0.71] & [0.01] \\
        \hline
        Year fixed effect & Yes & Yes\\
        Industry fixed effect & No & Yes\\
        Firm fixed effect & Yes & No\\
        Observations & 27760 & 27760\\
        R-squared & 0.2944 & 0.3921\\
        \hline
        \hline
    \end{tabular}
    \end{adjustbox}
\caption{\textbf{Determinants of firm-level cyber risk}}\bigskip
\footnotesize{This table reports the results of cyber risk regressions on firm characteristics. We include in turn year-, industry-, and firm-fixed effects. We report t-statistics in brackets. The variables are standardized, and the standard errors are clustered at the firm level. $^{*}$, $^{**}$, and $^{***}$ indicate significance at the 10\%, 5\% and 1\% levels, respectively. We define the characteristics in Table \ref{tab:variable_descriptions}.}
\label{tab:determinants_cyber_security}
\end{table}

\clearpage

\clearpage
\begin{table}
    \noindent\makebox[\textwidth]{%
    \begin{tabular}{lcccccc}
        & \multicolumn{5}{c}{ Value Weighted Portfolios} & \\
        \cmidrule(lr){2-6}
        & L & & & & H & H-L\\
        & P1 & P2 & P3 & P4 & P5 & P5-P1 \\
        \midrule
        \multicolumn{6}{l}{A. Portfolios sorted by cyber risk}\\
        Average excess returns
        & \textbf{0.88}$^{***}$ & \textbf{1.02}$^{***}$ & \textbf{1.13}$^{***}$ & \textbf{1.20}$^{***}$ & \textbf{1.44}$^{***}$ & \textbf{0.56}$^{*}$\\
        & [3.20] & [3.88] & [3.73] & [4.65] & [4.19] & [1.72] \\ 
        CAPM alpha
        & \textbf{-0.22} & \textbf{-0.06} & \textbf{-0.04} & \textbf{0.07} & \textbf{0.31} & \textbf{0.54}\\
        & [-0.95] & [-0.42] & [-0.35] & [1.21] & [1.61] & [1.32] \\
        FFC alpha
        & \textbf{-0.14} & \textbf{-0.01} & \textbf{0.03} &\textbf{0.04} & \textbf{0.24}$^{*}$ & \textbf{0.38}$^{*}$\\
        & [-1.20] & [-0.15] & [0.41] & [0.62] & [1.93] & [1.87] \\
        FF5 alpha
        & \textbf{-0.16} & \textbf{-0.08} & \textbf{0.03} & \textbf{0.04} & \textbf{0.25}$^{*}$ & \textbf{0.41}$^{**}$\\
        & [-1.62] & [-0.89] & [0.39] & [0.54] & [1.89] & [2.21] \\
        \\
        \multicolumn{6}{l}{B. Characteristics}\\
        Number of firms & 615.7 & 615.1 & 615.1 & 615.1 & 615.5 & -\\
        Cyber risk & 0.493 & 0.507 & 0.518 & 0.532 & 0.572 & -\\
        Sharpe Ratio & 0.637 & 0.775 & 0.801 & 0.892 & 1.037 & 0.628 \\
        Treynor Ratio & 0.031 & 0.037 & 0.038 & 0.042 & 0.050 & 3.283 \\
        Sortino Ratio & 0.942 & 1.199 & 1.263 & 1.462 & 1.790 & 2.733 \\
        \bottomrule
    \end{tabular}}
    \caption{\textbf{Average monthly excess returns and alphas}}\bigskip
    \footnotesize{This table reports the excess returns and alphas (in percent) of quintile portfolios of firms sorted on our cyber risk measure. \quotes{FFC} refers to the four-factor model of \cite{Carhart1997}, and \quotes{FF5} refers to the five-factor model of \cite{FamaFrench2015}. Panel B shows the average number of firms in each portfolio, the average cyber risk, and the annualized Sharpe, Treynor, and Sortino ratios. Newey-West (\citealp{NeweyWest1994}) t-statistics are reported in brackets. $^{*}$, $^{**}$, and $^{***}$ indicate significance at the 10\%, 5\% and 1\% levels, respectively. The period is from January 2009 to December 2022.}
    \label{tab:excess_returns_alphas_full_sample}
\end{table}

\clearpage
\begin{table}
    \noindent\makebox[\textwidth]{%
    \begin{tabular}{lccccc}
        & \multicolumn{5}{c}{Value Weighted Portfolios}\\
        \cmidrule(lr){2-6}
        & Cyber Q1 (Low) & Cyber Q2 & Cyber Q3 & Cyber Q4 & Cyber Q5 (High)\\
        \midrule
        A. sorted on market beta\\
        Beta Q1 (Low) & 0.94 & 1.01 & 0.85 & 1.09 & 0.91 \\
        Beta Q2 & 1.04 & 1.03 & 1.23 & 1.25 & 1.55 \\
        Beta Q3 & 0.98 & 1.22 & 1.25 & 1.36 & 1.39 \\
        Beta Q4 & 1.36 & 1.19 & 1.38 & 1.29 & 1.96 \\
        Beta Q5 (High) & 1.61 & 1.71 & 1.79 & 1.86 & 1.72 \\
        \midrule
        B. sorted on book-to-market\\
        BM Q1 (Low) & 0.90 & 1.09 & 1.22 & 1.28 & 1.54 \\
        BM Q2 & 1.02 & 1.09 & 1.32 & 1.23 & 1.36 \\
        BM Q3 & 0.99 & 1.04 & 1.23 & 1.18 & 1.51 \\
        BM Q4 & 1.04 & 1.17 & 1.29 & 1.25 & 1.37 \\
        BM Q5 (High) & 1.87 & 1.77 & 1.73 & 1.64 & 1.71 \\
        \midrule
        C. sorted on size \\
        Size Q1 (Small) & 1.33 & 1.25 & 1.45 & 1.77 & 1.24 \\
        Size Q2 & 1.35 & 1.54 & 1.23 & 1.18 & 1.38 \\
        Size Q3 & 1.43 & 1.39 & 1.19 & 1.33 & 1.47 \\
        Size Q4 & 1.16 & 1.21 & 1.29 & 1.24 & 1.30\\
        Size Q5 (Large) & 0.86 & 1.00 & 1.14 & 1.25 & 1.52\\
        \bottomrule
    \end{tabular}}
    \caption{\textbf{Double sorted portfolios}}\bigskip
    \footnotesize{This table reports average monthly excess returns in percent of double-sorted portfolios, first on market beta, book-to-market, or size and then on cyber risk. The period is from January 2009 to December 2022.}
    \label{tab:double_sorts}
\end{table}

\clearpage
\begin{table}
\vspace{-2em}
\begin{adjustbox}{width=0.77\textwidth,center}
    \noindent\makebox[\textwidth]{%
    \begin{tabular}{llllllll}
    \toprule
    \toprule
    \multicolumn{8}{c}{Portfolio returns $R_{P, t+1}$}\\
    \midrule
    & \multicolumn{1}{c}{(1)} & \multicolumn{1}{c}{(2)} & \multicolumn{1}{c}{(3)} & \multicolumn{1}{c}{(4)} & \multicolumn{1}{c}{(5)} & \multicolumn{1}{c}{(6)} & \multicolumn{1}{c}{(7)}\\
    \midrule
    $\beta_{Market, t}$ & \textbf{-0.005} & & \textbf{-0.025} & \textbf{0.065}  & \textbf{0.024} & & \textbf{-0.033}\\
    & [-0.064] & & [-0.429] & [0.997] & [0.275] & & [-0.578]\\
    Cyber risk$_t$ & & \textbf{0.183}$^{*}$& \textbf{0.182}$^{**}$ & \textbf{0.183}$^{*}$ & \textbf{0.172}$^{**}$ & \textbf{0.150}$^{*}$ & \textbf{0.176}$^{**}$\\
    & & [1.794] & [1.994] & [1.913] & [2.037] & [1.736] & [2.103]\\
    $\beta_{HML, t}$ & & & & \textbf{0.027} & \textbf{-0.012} &  & \textbf{-0.033}\\
    & & & & [0.439] & [-0.126] &  & [-0.293]\\
    $\beta_{SMB, t}$ & & & & \textbf{0.069}  & \textbf{0.049} &  & \textbf{-0.042}\\
    & & & & [0.835] & [0.536] &  & [-0.385]\\
    $\beta_{MOM, t}$ & & & & \textbf{0.011} & & &\\
    & & & & [0.153] & &\\
    $\beta_{RMW, t}$ & & & & & \textbf{-0.085} & & \textbf{-0.086}\\
    & & & & & [-1.436] & & [-1.150]\\
    $\beta_{CMA, t}$ & & & & & \textbf{-0.088} & & \textbf{0.011}\\
    & & & & & [-0.816] & & [0.076]\\
    \\
    $\beta_{Consumer Durables, t}$ & & & & & & \textbf{-0.005}\\
    & & & & & & [-0.053]\\
    $\beta_{Energy, t}$ & & & & & & \textbf{-0.005}\\
    & & & & & & [-0.064]\\
    $\beta_{Chemicals, t}$ & & & & & & \textbf{0.025}\\
    & & & & & & [0.351]\\
    $\beta_{Telecommunications, t}$ & & & & & & \textbf{0.045}\\
    & & & & & & [0.619]\\
    $\beta_{Retail, t}$ & & & & & & \textbf{0.116}\\
    & & & & & & [1.015]\\
    $\beta_{Healthcare, t}$ & & & & & & \textbf{0.048}\\
    & & & & & & [0.623]\\
    $\beta_{Finance, t}$ & & & & & & \textbf{0.101}\\
    & & & & & & [1.104]\\
    Constant & \textbf{1.445}$^{***}$ & \textbf{1.465}$^{***}$ & \textbf{1.457}$^{***}$ & \textbf{1.455}$^{***}$ & \textbf{1.476}$^{***}$ & \textbf{0.945}$^{**}$ & \textbf{1.362}$^{***}$\\
    & [5.311] & [5.540] & [5.493]& [5.413] & [5.450] & [2.525] & [5.015]\\
    \midrule
    $\overline{R2_{adj}}$ & 0.007 & 0.134 & 0.186 & 0.258 & 0.284 & 0.302 & 0.295\\
    MAPE & 1.360 & 1.312 & 1.233 & 1.064 & 0.987 & 0.914 & 1.243\\
    Nb portfolios & 20 & 20 & 20 & 20 & 20 & 20 & 31 \\
    \hline
    \hline
    \end{tabular}}
    \end{adjustbox}
    \caption{\textbf{Fama-MacBeth regressions}}\bigskip
    \footnotesize{This table reports the results of Fama-MacBeth regressions of 20 value-weighted portfolios arithmetic returns sorted on their cyber risk. In Specification (7), we extend the set of test assets with an additional 11 industry portfolios. These portfolios are regressed each month on portfolio value-weighted betas with the market, HML, SMB, MOM, RMW, and CMA and on portfolio value-weighted betas with industry portfolios in Specification (6). \quotes{Cyber risk} is the value-weighted cyber-risk of each portfolio. The betas are standardized before the second step-regressions and estimated with a two-year rolling window. HML and SMB refer to the book-to-market and size factors from \cite{FamaFrench1992}. MOM refers to the momentum factor from \cite{Carhart1997}. CMA and RMW refer to the investment and operating profitability factors from \cite{FamaFrench2015}. $\overline{R2_{adj}}$ is the average adjusted R-squared, and MAPE is the mean average pricing error. Newey-West t-statistics are reported in brackets. $^{*}$, $^{**}$, and $^{***}$ indicate significance at the 10\%, 5\% and 1\% levels, respectively. The period is from January 2009 to December 2022.}
    \label{tab:Fama_MacBeth}
\end{table}

\clearpage
\begin{table}
    \noindent\makebox[\textwidth]{%
    \begin{tabular}{lcccccc}
    \toprule
    \toprule
    & GRS & p-value & $\overline{R2}$ & GRS & p-value & $\overline{R2}$\\
    \midrule
    & \multicolumn{3}{c}{Sorted on cyber risk}
    & \multicolumn{3}{c}{Sorted on market beta}\\
    \cmidrule(lr){2-4}
    \cmidrule(lr){5-7}
    FF5 & 1.211 & 0.253 & 0.869 & 0.712 & 0.802 & 0.783\\
    FF5 + CyberFactor & 0.947 & 0.530 & 0.886 & 0.825 & 0.680 & 0.801\\
    \\
    & \multicolumn{3}{c}{Sorted on size}
    & \multicolumn{3}{c}{Sorted on book-to-market}\\
    \cmidrule(lr){2-4}
    \cmidrule(lr){5-7}
    FF5 & 1.490 & 0.093 & 0.879 & 1.709 & 0.038 & 0.878\\
    FF5 + CyberFactor & 1.458 & 0.106 & 0.880 & 1.417 & 0.124 & 0.883\\
    \bottomrule
    \bottomrule
    \end{tabular}}
    \caption{\textbf{GRS test statistics}}\bigskip
    \footnotesize{This table reports the results of time series regressions of 20 value-weighted portfolios sorted on cyber-risk on the five-factor model of \cite{FamaFrench2015} (FF5) and on the \quotes{CyberFactor}, \textit{i.e.} the factor built as the long-short of extreme quintile portfolios sorted on our cyber risk score. The study period is from January 2009 to December 2022.}
    \label{tab:GRS}
\end{table}

\clearpage

\clearpage

\clearpage
\begin{table}
    \begin{adjustbox}{width=1.1\textwidth,center}
    \noindent\makebox[\textwidth]{%
    \begin{tabular}{lcccc}
    \toprule
    \toprule
     Prior Multiple & 1.25 & 1.5 & 2 & 3\\
    \midrule
    Mkt HML RMW CMA CyberFactor & 19.40 & 21.18 & 21.31 & 18.79\\
    Mkt RMW CMA CyberFactor & 16.54 & 19.01 & 21.89 & 26.08\\
    Mkt SMB HML RMW CMA CyberFactor & 18.38 & 18.38 & 15.92 & 10.41\\
    Mkt SMB RMW CMA CyberFactor & 11.92 & 12.85 & 12.81 & 11.24\\
    Mkt CMA CyberFactor & 5.76 & 5.87 & 7.15 & 11.18\\
    \bottomrule
    \bottomrule
    \end{tabular}
    }
    \end{adjustbox}
    \caption{\textbf{Prior sensitivity of the posterior model probabilities}}\bigskip
    \footnotesize{This table reports the posterior model probabilities (in percent) for the top five models (ranked at the end of the sample) for different values of the prior multiple. The top five models remain the same for every prior multiple. \quotes{Mkt} refers to the excess return on the market from the Kenneth French data repository. HML and SMB refer to the book-to-market and size factors of \cite{FamaFrench1992}. CMA and RMW refer to the investment and operating profitability factors of \cite{FamaFrench2015}. \quotes{CyberFactor} refers to the long-short portfolio built on our cyber risk score. The study period is from January 2010 to December 2022.}
    \label{tab:prior_sensitivity}
\end{table}

\clearpage
\begin{table}
    \noindent\makebox[\textwidth]{%
    \begin{tabular}{lcccc}
        & \multicolumn{3}{c}{Value Weighted Portfolios} & \\
        \cmidrule(lr){2-4}
        & L & & H & H-L\\
        & P1 & P2 & P3 & P3-P1 \\
        \midrule
        \multicolumn{1}{l}{A. Portfolios sorted by cyber risk}\\
        Average excess returns
        & \textbf{0.71} & \textbf{0.88}$^{**}$ & \textbf{1.24}$^{***}$ & \textbf{0.53}$^{**}$\\
        & [1.83] & [2.25] & [3.37] & [2.00] \\ 
        CAPM alpha
        & \textbf{-0.58}$^{***}$ & \textbf{-0.35} & \textbf{-0.07} & \textbf{0.51}$^*$\\
        & [-2.75] & [-1.11] & [-0.35] & [1.73] \\
        FFC alpha
        & \textbf{-0.40}$^*$ & \textbf{-0.23} & \textbf{-0.05} &\textbf{0.35} \\
        & [-1.89] & [-0.82] & [-0.37] & [1.38]\\
        FF5 alpha
        & \textbf{-0.37}$^*$ & \textbf{-0.27} & \textbf{0.01} & \textbf{0.38} \\
        & [-1.73] & [-0.97] & [0.09] & [1.54] \\
        \multicolumn{1}{l}{B. Characteristics} \\
        Number of firms & 248.5 & 247.9 & 248.3 & -\\
        Cyber risk & 0.486 & 0.504 & 0.534 & -\\
        Sharpe Ratio & 0.489 & 0.645 & 0.856 & 0.567\\
        Treynor Ratio & 0.023 & 0.030 & 0.039 & 3.976\\
        Sortino Ratio & 0.735 & 1.071 & 1.463 & 2.349\\
        \bottomrule
    \end{tabular}}
    \caption{\textbf{Average monthly excess returns and alphas of firms classified as zero-risk by \cite{FlorackisLoucaMichaelyWeber2023}}}\bigskip
    \footnotesize{This table reports the average monthly excess returns and alphas (in percent) of firms classified as zero-risk when using the procedure of \cite{FlorackisLoucaMichaelyWeber2023}. FFC refers to the four-factor model of \cite{Carhart1997}, and FF5 refers to the five-factor model of \cite{FamaFrench2015}. Panel B reports the average number of firms in each portfolio, the average cyber risk, and the annualized Sharpe, Treynor, and Sortino ratios. Newey-West (\citealp{NeweyWest1994}) t-statistics are reported in brackets. $^{*}$, $^{**}$, and $^{***}$ indicate significance at the 10\%, 5\% and 1\% levels, respectively. The period is from January 2009 to December 2018.}
    \label{tab:zero_risk_firm_sorts}
\end{table}

\clearpage
\begin{table}
    \noindent\makebox[\textwidth]{%
    \begin{tabular}{lcccccc}
        & \multicolumn{5}{c}{ Value Weighted Portfolios} & \\
        \cmidrule(lr){2-6}
        & L & & & & H & H-L\\
        & P1 & P2 & P3 & P4 & P5 & P5-P1 \\
        \midrule
        \multicolumn{6}{l}{A. Portfolios sorted by cyber risk}\\
        Average excess returns
        & \textbf{0.97}$^{***}$ & \textbf{1.05}$^{***}$ & \textbf{1.15}$^{***}$ & \textbf{1.19}$^{***}$ & \textbf{1.48}$^{***}$ & \textbf{0.51}\\
        & [3.61] & [4.01] & [3.99] & [4.10] & [4.24] & [1.62] \\ 
        CAPM alpha
        & \textbf{-0.15} & \textbf{0.00} & \textbf{-0.01} & \textbf{0.07} & \textbf{0.33} & \textbf{0.48}\\
        & [-0.81] & [-0.02] & [-0.04] & [0.85] & [1.54] & [1.26] \\
        FFC alpha
        & \textbf{-0.09} & \textbf{0.05} & \textbf{0.04} &\textbf{0.08} & \textbf{0.24}$^{*}$ & \textbf{0.33}$^{*}$\\
        & [-0.98] & [0.46] & [0.45] & [1.16] & [1.82] & [1.69] \\
        FF5 alpha
        & \textbf{-0.10} & \textbf{-0.03} & \textbf{0.05} & \textbf{0.05} & \textbf{0.25}$^{*}$ & \textbf{0.35}$^{**}$\\
        & [-1.40] & [-0.38] & [0.51] & [85] & [1.90] & [2.10] \\
        \\
        \multicolumn{6}{l}{B. Characteristics}\\
        Number of firms & 592.9 & 592.3 & 592.1 & 592.1 & 592.6 & -\\
        Cyber risk & 0.482 & 0.496 & 0.507 & 0.521 & 0.565 & -\\
        Sharpe Ratio & 0.702 & 0.820 & 0.829 & 0.883 & 1.029 & 0.571 \\
        Treynor Ratio & 0.034 & 0.039 & 0.039 & 0.042 & 0.051 & 2.027 \\
        Sortino Ratio & 1.058 & 1.305 & 1.346 & 1.417 & 1.777 & 2.818 \\
        \bottomrule
    \end{tabular}}
    \caption{\textbf{Average monthly excess returns and alphas omitting Items 1A}}\bigskip
    \footnotesize{This table reports the average monthly excess returns and alphas (in percent) when items 1A from 10-K statements are omitted. FFC refers to the four-factor model of \cite{Carhart1997}, and FF5 refers to the five-factor model of \cite{FamaFrench2015}. Panel B reports the average number of firms in each portfolio, the average cyber risk, and the annualized Sharpe, Treynor, and Sortino ratios. Newey-West (\citealp{NeweyWest1994}) t-statistics are reported in brackets. $^{*}$, $^{**}$, and $^{***}$ indicate significance at the 10\%, 5\% and 1\% levels, respectively. The period is from January 2009 to December 2022.}
    \label{tab:excess_returns_alphas_no_1A}
\end{table}

\clearpage
\begin{table}
    \noindent\makebox[\textwidth]{%
    \begin{tabular}{lcccccc}
        & \multicolumn{5}{c}{Value Weighted Portfolios} & \\
        \cmidrule(lr){2-6}
        & L & & & & H & H-L\\
        & P1 & P2 & P3 & P4 &  P5 & P5-P1 \\
        \midrule
        \multicolumn{6}{l}{A. Portfolios sorted by the long-run cyber risk}\\
        Average excess returns
        & \textbf{0.86}$^{***}$ & \textbf{1.13}$^{***}$ & \textbf{1.14}$^{***}$ & \textbf{1.18}$^{***}$ & \textbf{1.45}$^{***}$ & \textbf{0.60}$^{*}$\\
        & [3.00]& [3.94] & [3.93] & [4.35] & [4.23] & [1.70] \\ 
        CAPM alpha
        & \textbf{-0.26} & \textbf{-0.01} & \textbf{-0.03} & \textbf{0.11}$^{*}$ & \textbf{0.31} & \textbf{0.57}\\
        & [-1.05] & [-0.04] & [-0.27] & [1.71] & [1.52] & [1.33] \\
        FFC alpha
        & \textbf{-0.17} & \textbf{0.01} & \textbf{0.05} &\textbf{ 0.10} & \textbf{0.23}$^{*}$ & \textbf{0.40}$^{**}$\\
        & [-1.56] & [0.67] & [0.75] & [1.24] & [1.83] & [2.01] \\
        FF5 alpha
        & \textbf{-0.17}$^{*}$& \textbf{0.03} & \textbf{-0.01} & \textbf{0.06} & \textbf{0.25}$^{*}$ & \textbf{0.43}$^{**}$\\
        & [-1.80] & [0.34] & [-0.16] & [0.93] & [1.94] & [2.28] \\
        \\
        \multicolumn{6}{l}{B. Characteristics}\\
        Number of firms & 615.7 & 615.1 & 615.1 & 615.1 & 615.5 & -\\
        Long-run cyber risk & 0.490 & 0.501 & 0.511 & 0.524 & 0.567 & -\\
        Sharpe Ratio & 0.610 & 0.818 & 0.816 & 0.918 & 1.027 & 0.638 \\
        Treynor Ratio & 0.030 & 0.039 & 0.038 & 0.044 & 0.050 & 2.572 \\
        Sortino Ratio & 0.893 & 1.305 & 1.294 & 1.525 & 1.768 & 2.643 \\
        \bottomrule
    \end{tabular}}
    \caption{\textbf{Average monthly excess returns and alphas using the long-run cyber risk}}\bigskip 
    \footnotesize{This table reports the average monthly excess returns and alphas (in percent) when the portfolios are sorted on our long-run cyber risk measure, \textit{i.e.} the cumulative cyber risk for each firm over the period. FFC refers to the four-factor model of \cite{Carhart1997}, and FF5 refers to the five-factor model of \cite{FamaFrench2015}. Panel B shows the average number of firms in each portfolio, the average long-run cyber risk, and the annualized Sharpe, Treynor, and Sortino ratios. Newey-West t-statistics are reported in brackets. $^{*}$, $^{**}$, and $^{***}$ indicate significance at the 10\%, 5\% and 1\% levels, respectively. The period is from January 2009 to December 2022.}
    \label{tab:excess_return_alphas_long_run}
\end{table}

\clearpage
\begin{table}
    \noindent\makebox[\textwidth]{%
    \begin{tabular}{lcccccc}
        & \multicolumn{5}{c}{ Value Weighted Portfolios} & \\
        \cmidrule(lr){2-6}
        & L & & & & H & H-L\\
        & P1 & P2 & P3 & P4 &  P5 & P5-P1 \\
        \midrule
        \multicolumn{6}{l}{A. Portfolios sorted by cyber risk}\\
        Average excess returns
        & \textbf{0.88}$^{***}$ & \textbf{1.02}$^{***}$ & \textbf{1.13}$^{***}$ & \textbf{1.22}$^{***}$ & \textbf{1.45}$^{***}$ & \textbf{0.57}$^{*}$\\
        & [3.19]& [3.86] & [3.71] & [4.73] & [4.18] & [1.74] \\ 
        CAPM alpha
        & \textbf{-0.22} & \textbf{-0.07} & \textbf{-0.05} & \textbf{0.08} & \textbf{0.33} & \textbf{0.55}\\
        & [-0.94] & [-0.44] & [-0.39] & [1.38] & [1.67] & [1.34] \\
        FFC alpha
        & \textbf{-0.14} & \textbf{-0.02} & \textbf{0.03} &\textbf{0.05} & \textbf{0.25}$^{**}$ & \textbf{0.39}$^{*}$\\
        & [-1.18] & [-0.19] & [0.34] & [0.81] & [2.027] & [1.91] \\
        FF5 alpha
        & \textbf{-0.16} & \textbf{-0.08} & \textbf{0.03} & \textbf{0.06} & \textbf{0.26}$^{**}$ & \textbf{0.41}$^{**}$\\
        & [-1.61] & [-0.93] & [0.35] & [0.72] & [1.97] & [2.26] \\
        \\
        \multicolumn{6}{l}{B. Characteristics}\\
        Number of firms & 611.9 & 611.3 & 611.3 & 612.3 & 611.7 & -\\
        Cyber risk & 0.493 & 0.507 & 0.518 & 0.532 & 0.571 & -\\
        Sharpe Ratio & 0.637 & 0.772 & 0.796 & 0.903 & 1.049 & 0.641 \\
        Treynor Ratio & 0.031 & 0.037 & 0.038 & 0.042 & 0.051 & 4.408 \\
        Sortino Ratio & 0.942 & 1.196 & 1.253 & 1.483 & 1.812 & 2.759 \\
        \bottomrule
    \end{tabular}}
    \caption{\textbf{Average monthly excess returns and alphas excluding cybersecurity firms}}\bigskip
    \footnotesize{This table reports the average monthly excess returns and alphas (in percent) when we exclude cybersecurity firms from the sample. We identify cybersecurity firms from the \quotes{HACK} cybersecurity ETF. FFC refers to the four-factor model of \cite{Carhart1997}, and FF5 refers to the five-factor model of \cite{FamaFrench2015}. Panel B shows the average number of firms in each portfolio, the average cyber risk, and the annualized Sharpe, Treynor, and Sortino ratios. Newey-West t-statistics are reported in brackets. $^{*}$, $^{**}$, and $^{***}$ indicate significance at the 10\%, 5\% and 1\% levels, respectively. The period is from January 2009 to December 2022.}
    \label{tab:excess_return_alphas_no_cyber}
\end{table}
\clearpage

\clearpage
\begin{table}
    \noindent\makebox[\textwidth]{%
    \begin{tabular}{lcccccc}
        & \multicolumn{5}{c}{Value Weighted Portfolios} & \\
        \cmidrule(lr){2-6}
        & L & & & & H & H-L\\
        & P1 & P2 & P3 & P4 &  P5 & P5-P1 \\
        \midrule
        \multicolumn{6}{l}{A. Portfolios sorted by cyber risk}\\
        Average excess returns
        & \textbf{0.70}$^{**}$ & \textbf{0.97}$^{***}$ & \textbf{1.10}$^{***}$ & \textbf{1.23}$^{***}$ & \textbf{1.64}$^{***}$ & \textbf{0.94}$^{***}$\\
        & [2.41]& [3.58] & [3.69] & [5.76] & [5.76] & [3.23] \\ 
        CAPM alpha
        & \textbf{-0.52}$^{***}$ & \textbf{-0.20} & \textbf{-0.14} & \textbf{0.06} & \textbf{0.51}$^{**}$ & \textbf{1.03}$^{***}$\\
        & [-3.11] & [-1.50] & [-1.07] & [0.95] & [2.40] & [2.95] \\
        FFC alpha
        & \textbf{-0.28}$^{***}$ & \textbf{-0.06} & \textbf{0.00} &\textbf{-0.03} & \textbf{0.27}$^{*}$ & \textbf{0.55}$^{***}$\\
        & [-3.58] & [-0.73] & [0.01] & [-0.52] & [1.91] & [2.83] \\
        FF5 alpha
        & \textbf{-0.26}$^{***}$& \textbf{-0.08} & \textbf{0.03} & \textbf{-0.06} & \textbf{0.30}$^{*}$ & \textbf{0.56}$^{***}$\\
        & [-3.45] & [-0.88] & [0.36] & [-0.95] & [1.90] & [3.01] \\
        \\
        \multicolumn{6}{l}{B. Characteristics}\\
        Number of firms & 600.4 & 599.9 & 599.9 & 599.9 & 600.2 & -\\
        Cyber risk & 0.490 & 0.504 & 0.515 & 0.529 & 0.570 & -\\
        Sharpe Ratio & 0.511 & 0.752 & 0.807 & 0.965 & 1.272 & 1.157 \\
        Treynor Ratio & 0.024 & 0.034 & 0.037 & 0.043 & 0.060 & -0.714 \\ 
        Sortino Ratio & 0.723 & 1.137 & 1.257 & 1.571 & 2.393 & 4.319 \\
        \bottomrule
    \end{tabular}}
    
    \caption{\textbf{Average monthly excess returns and alphas before the first release of \cite{FlorackisLoucaMichaelyWeber2023}.}}\bigskip\footnotesize This table reports the average monthly excess returns and alphas (in percent) before the first release of \cite{FlorackisLoucaMichaelyWeber2023} on SSRN in November 2020. FFC refers to the four-factor model of \cite{Carhart1997}, and FF5 refers to the five-factor model of \cite{FamaFrench2015}. Panel B reports the average number of firms in each portfolio, the average cyber risk, and the annualized Sharpe, Treynor, and Sortino ratios.  Newey-West t-statistics are reported in brackets. $^{*}$, $^{**}$, and $^{***}$ indicate significance at the 10\%, 5\% and 1\% levels, respectively. The period is from January 2009 to October 2020.\\
    \label{tab:excess_return_alphas_before}
\end{table}

\clearpage
\begin{table}
    \noindent\makebox[\textwidth]{%
    \begin{tabular}{lcccccc}
        & \multicolumn{5}{c}{Value Weighted Portfolios} & \\
        \cmidrule(lr){2-6}
        & L & & & & H & H-L\\
        & P1 & P2 & P3 & P4 &  P5 & P5-P1 \\
        \midrule
        \multicolumn{6}{l}{A. Portfolios sorted by cyber risk}\\
        Average excess returns
        & \textbf{2.16}$^{**}$ & \textbf{1.56} & \textbf{1.34} & \textbf{1.55} & \textbf{0.67} & \textbf{-1.49}$^{**}$\\
        & [2.18] & [1.64] & [0.97] & [1.19] & [0.44] & [-2.11] \\ 
        CAPM alpha
        & \textbf{1.34}$^{***}$ & \textbf{0.71}$^{***}$ & \textbf{0.34} & \textbf{0.54}$^{***}$ & \textbf{-0.40}$^{*}$ & \textbf{-1.74}$^{***}$\\
        & [4.37] & [2.91] & [1.15] & [3.44] & [-1.67] & [-3.35] \\
        FFC alpha
        & \textbf{0.57}$^{*}$ & \textbf{0.19} & \textbf{-0.07} & \textbf{0.54}$^{***}$ & \textbf{0.17} & \textbf{-0.39}\\
        & [1.76] & [0.71] & [-0.37] & [3.10] & [0.61] & [-0.67] \\
        FF5 alpha
        & \textbf{0.34}& \textbf{-0.15} & \textbf{-0.15} & \textbf{0.65}$^{***}$ & \textbf{0.11} & \textbf{-0.23}\\
        & [0.89] & [-0.53] & [-0.74] & [3.63] & [0.38] & [-0.34] \\
        \\
        \multicolumn{6}{l}{B. Characteristics}\\
        Number of firms & 695.5 & 695.0 & 694.9 & 695.0 & 695.2 & -\\
        Cyber risk & 0.504 & 0.523 & 0.536 & 0.550 & 0.582 & -\\
        Sharpe Ratio & 1.374 & 1.009 & 0.758 & 0.873 & 0.357 & -1.287 \\
        Treynor Ratio & 0.088 & 0.061 & 0.045 & 0.051 & 0.021 & 0.092 \\
        Sortino Ratio & 2.798 & 1.830 & 1.280 & 1.665 & 0.529 & 0.640 \\
        \bottomrule
    \end{tabular}}
    \caption{\textbf{Average monthly excess returns and alphas after the first release of \cite{FlorackisLoucaMichaelyWeber2023}.}}
    \bigskip\footnotesize{This table reports the average monthly excess returns and alphas (in percent) after the first release of \cite{FlorackisLoucaMichaelyWeber2023} on SSRN in November 2020. FFC refers to the four-factor model of \cite{Carhart1997}, and FF5 refers to the five-factor model of \cite{FamaFrench2015}. Panel B reports the average number of firms in each portfolio, the average cyber risk, and the annualized Sharpe, Treynor, and Sortino ratios. Newey-West t-statistics are reported in brackets. $^{*}$, $^{**}$, and $^{***}$ indicate significance at the 10\%, 5\% and 1\% levels, respectively. The period is from November 2020 to December 2022.}
    \label{tab:excess_return_alphas_after}
\end{table}

\clearpage
\section*{Appendix}

\setcounter{table}{0}
\renewcommand{\thetable}{A\arabic{table}}
\setcounter{figure}{0}
\renewcommand{\thefigure}{A\arabic{figure}}

\begin{table}[h]
\begin{adjustbox}{width=1.15\textwidth,center}
    \begin{tabular}{lcc}
    \hline
    \hline
     Variable & Description & Source \\
    \hline
    Firm size (ln) & ln(total assets [at])  & Compustat\\
    Firm Age (ln) & ln(years) since the firm first appeared in Compustat & Compustat\\
    Book to market ratio & Common equity [ceq] / market equity [prc*shrout] & Compustat and CRSP\\
    Tobin's Q & (Total assets - common equity + market equity) / total assets & Compustat and CRSP \\
    ROA & Net income [ni] / total assets & Compustat\\
    Market Beta & 5-year rolling market beta [beta] & Compustat\\
    Intangible/Assets & Intangible assets [intan] / total assets & Compustat\\
    Debt/assets & Total Debt / Total Assets [debt\_assets] & WRDS Financial Ratios \\
    ROE & Net Income / Book Equity [roe] & WRDS Financial Ratios \\
    Price/Earnings & Stock Price / Earnings [pe\_exi] & WRDS Financial Ratios\\
    Profit Margin & Gross Profit / Sales [gpm] & WRDS Financial Ratios \\
    Asset Turnover & Sales / Total Assets [at\_turn] & WRDS Financial Ratios\\
    Cash Ratio & (Cash + Short-term Investments) / Current Liabilities [cash\_ratio] & WRDS Financial Ratios\\
    Sales/Invested Capital & Sales per dollar of Invested Capital [sale\_invcap] &  WRDS Financial Ratios\\
    Capitalization Ratio &  Long-term Debt / (Long-term Debt + Equity) [capital\_ratio] & WRDS Financial Ratios \\
    R\&D/Sales & R\&D expenses / Sales [RD\_SALE] & WRDS Financial Ratios \\
    ROCE & Earnings Before Interest and Taxes / average Capital Employed [roce] & WRDS Financial Ratios\\
    \hline
    \hline
    \end{tabular}
    \end{adjustbox}
    \caption{\textbf{Variable definitions}}\bigskip
    \footnotesize{This table reports the variable names used throughout the paper, their description, and their source. Square brackets indicate variable name definitions in CRSP and Compustat.}
\label{tab:variable_descriptions}
\end{table}

\clearpage
\begin{table}
\noindent\makebox[\textwidth]{%
    \begin{tabular}{cp{115mm}cc}
        \toprule
        \toprule
        Score & \multicolumn{1}{c}{Preprocessed paragraph} & Ticker & Tactic\\
        \hline
        \footnotesize 0.593 & \footnotesize currently available internet browsers allow users modify browser settings remove cookies prevent cookies stored hard drives however third persons able penetrate network security gain access otherwise misappropriate users personal information subject liability liability include claims misuses personal information unauthorized marketing purposes unauthorized use credit cards & \footnotesize VSTY & \footnotesize Defense Evasion\\
        \footnotesize 0.590 & \footnotesize network security data recovery measures may adequate protect computer viruses break ins similar disruptions unauthorized tampering computer systems theft sabotage type security breach respect proprietary confidential information electronically stored including research clinical data material adverse impact business operating results financial condition & \footnotesize LXRX & \footnotesize Collection \\
        \footnotesize 0.583 & \footnotesize domain names derive value individual ability remember names therefore assurance domain name lose value example users begin rely mechanisms domain names access online resources government regulation internet regulation increasing number laws regulations pertaining internet & \footnotesize VSTY & \footnotesize Credential Access\\
        \footnotesize 0.577 & \footnotesize perceived actual unauthorized disclosure information collect breach security harm business factors beyond control cause interruptions operations may adversely affect reputation marketplace business financial condition results operations timely development implementation continuous uninterrupted performance hardware network applications internet systems including may provided third parties important facets delivery products services customers & \footnotesize MDAS & \footnotesize Credential Access\\
        \footnotesize 0.571 & \footnotesize unauthorized parties may attempt copy aspects products obtain use information regard proprietary others may independently develop otherwise acquire similar competing technologies methods design around patents cases rely trade secret laws confidentiality agreements protect confidential proprietary information processes technology & \footnotesize CSCD & \footnotesize Collection\\
        \footnotesize 0.570 & \footnotesize possible cookies may become subject laws limiting prohibiting use term cookies refers information keyed specific server file pathway directory location stored user hard drive possibly without user knowledge used among things track demographic information target advertising & \footnotesize VSTY & \footnotesize Discovery\\
        \footnotesize 0.561 & \footnotesize cannot certain advances computer capabilities discoveries field cryptography developments result compromise breach algorithms use protect content transactions website proprietary information databases anyone able circumvent security measures misappropriate proprietary confidential customer company information cause interruptions operations & \footnotesize VSTY & \footnotesize Impact\\
        \footnotesize 0.558 & \footnotesize ordering delivery customers ready place order proceed shopping cart function directly checkout page orders placed online website via toll free telephone number customer service agents available take orders customers access internet uncomfortable placing order online & \footnotesize VSTY & \footnotesize Credential Access\\
        \bottomrule
        \bottomrule
    \end{tabular}}

\caption{\textbf{Top scoring paragraphs from the doc2vec validation sample}}\bigskip
\footnotesize{This exhibit reports the paragraphs selected after preprocessing (as described in section \ref{subsection:preprocessing}). \quotes{Tactic} refers to the MITRE tactic the paragraph is most similar to, as measured by cosine similarity. The tickers of the ten companies in the validation sample are CSCD, GTS, LXRX, MDAS, PBY, PZZA, UMH, VALU, VSTY and VXRT.}
\label{tab:top_paragraphs_doc2vec_validation}
\end{table}

\clearpage
\begin{table}
    \begin{adjustbox}{width=\textwidth,center}
    \noindent\makebox[\textwidth]{%
    \begin{tabular}{lcp{145mm}}
    \toprule
    \toprule
     Ticker & Filling year & \multicolumn{1}{c}{Paragraph}\\
    \midrule
    \footnotesize STX & \footnotesize 2008 & 
    \footnotesize System failures caused by events beyond our control could adversely affect computer equipment and electronic data on which our operations depend. Our operations are dependent upon our ability to protect our computer equipment and the electronic data stored in our databases from damage by, among other things, earthquake, fire, natural disaster, power loss, telecommunications failures, unauthorized intrusion and other catastrophic events. As our operations become more automated and increasingly interdependent, our exposure to the risks posed by these types of events will increase. While we continue to improve our disaster recovery processes, system failures and other interruptions in our operations could have a material adverse effect on our business, results of operations and financial condition.\\
    \footnotesize MANH &  \footnotesize 2014 & \footnotesize Our software may contain undetected errors or “bugs” resulting in harm to our reputation which could adversely impact our business, results of operations, cash flow, and financial condition. Software products as complex as those offered by us might contain undetected errors or failures when first introduced or when new versions are released,. Despite testing, we cannot ensure that errors will not be found in new products or product enhancements after commercial release,. Any errors could cause substantial harm to our reputation, result in additional unplanned expenses to remedy any defects, delay the introduction of new products, result in the loss of existing or potential customers, or cause a loss in revenue. Further, such errors could subject us to claims from our customers for significant damages, and we cannot assure you that courts would enforce the provisions in our customer agreements that limit our liability for damages. In turn, our business, results of operations, cash flow, and financial condition could be materially adversely affected.\\
    \footnotesize MBCN & \footnotesize 2017 & \footnotesize Material breaches in security of bank systems may have a significant effect on the Company business. We collect, process and store sensitive consumer data by utilizing computer systems and telecommunications networks operated by both banks and third party service providers. We have security, backup and recovery systems in place, as well as a business continuity plan to ensure systems will not be inoperable. We also have security to prevent unauthorized access to the system. In addition, we require third party service providers to maintain similar controls. However, we cannot be certain that these measures will be successful. A security breach in the system and loss of confidential information could result in losing customers’ confidence and thus the loss of their business as well as additional significant costs for privacy monitoring activities.\\
    \bottomrule
    \bottomrule
    \end{tabular}}
    \end{adjustbox}
    \caption{\textbf{Examples of paragraphs missed by the dictionary approach of \cite{FlorackisLoucaMichaelyWeber2023}}}\bigskip
    \footnotesize{}
    \label{tab:missed_paragraphs}
\end{table}

\clearpage
\begin{table}
\vspace{-4em}
\noindent\makebox[\textwidth]{%
    \begin{tabular}{cp{115mm}cc}
        \toprule
        \toprule
        Score & \multicolumn{1}{c}{Paragraph} & Ticker & \makecell{Sub-technique \\ (Tactic)}\\
        \hline
        \footnotesize -0.097 & \footnotesize The increase in income from operations during 2021 was primarily the result of higher net sales and production volumes and improved margins, which benefited from positive pricing impacts and a favorable sales mix that offset substantial inflationary material and freight cost pressures along with higher manufacturing costs. Net income per diluted share was favorably impacted by the reversal of valuation allowances previously established against our deferred tax assets in both the United States and Brazil during 2021 & \footnotesize AGCO & \footnotesize \makecell{\\ SSH Authorized Keys\\ (Persistence)}\\
        \footnotesize -0.092 & \footnotesize In addition, we generally would expect to be able to recover a significant portion of the amounts paid under such guarantees from the sale of the underlying financed farm equipment, as the fair value of such equipment is expected to offset a substantial portion of the amounts paid. We also guarantee indebtedness owed to certain of our finance joint ventures if dealers or end users default on loans. & \footnotesize AGCO & \footnotesize \makecell{\\ Fast Flux DNS\\ (Command and Control)}\\
        \footnotesize -0.099 & \footnotesize In December 2021, the Company completed the sale of its Le Parfait brand in Europe and a previously closed plant in the Americas.  Gross proceeds on these divestitures were approximately \$113 million and the related pretax gains (including costs directly attributable to the sale) were approximately \$84 million (\$70 million after tax) in 2021.  The pretax gains were recorded to Other income (expense), net on the Consolidated Results of Operations. In January 2021, the Company completed the sale of its plant in Argentina.& \footnotesize OI & \footnotesize \makecell{\\ GUI Input Capture\\ (Collection)}\\
       \footnotesize -0.088 & \footnotesize In 2020, the Company recognized a net gain (including costs directly attributable to the sale of ANZ and subject to post-closing adjustments) on the divestiture of approximately \$275 million, which was reported on the Other income (expense), net line in the Consolidated Results of Operations.  In addition, at closing, certain subsidiaries of the Company entered into certain ancillary agreements with Visy and the ANZ businesses in respect of the provision of certain transitional and technical services to the ANZ businesses. & \footnotesize OI & \footnotesize \makecell{\\Scheduled Transfer \\ (Exfiltration)}\\
        \bottomrule
        \bottomrule
    \end{tabular}  
}
\caption{\textbf{Examples of negative scoring paragraphs}}\bigskip
\footnotesize{This exhibit reports examples of paragraphs with negative similarities, as measured by cosine similarity. \quotes{Sub-technique} refers to the MITRE sub-technique the paragraph is most dissimilar to. The displayed score is the cosine similarity between the paragraph and the previously mentioned sub-technique. Given their low cyber risk score, we select two firms, AGCO Corporation (AGCO) and O-I Glass Inc (OI). The 10-K statements are filed in 2022.}
\label{tab:negative_paragraphs}
\end{table}

\clearpage
\begin{figure} 
    \noindent\makebox[\textwidth]{%
    \includegraphics[width=\textwidth]{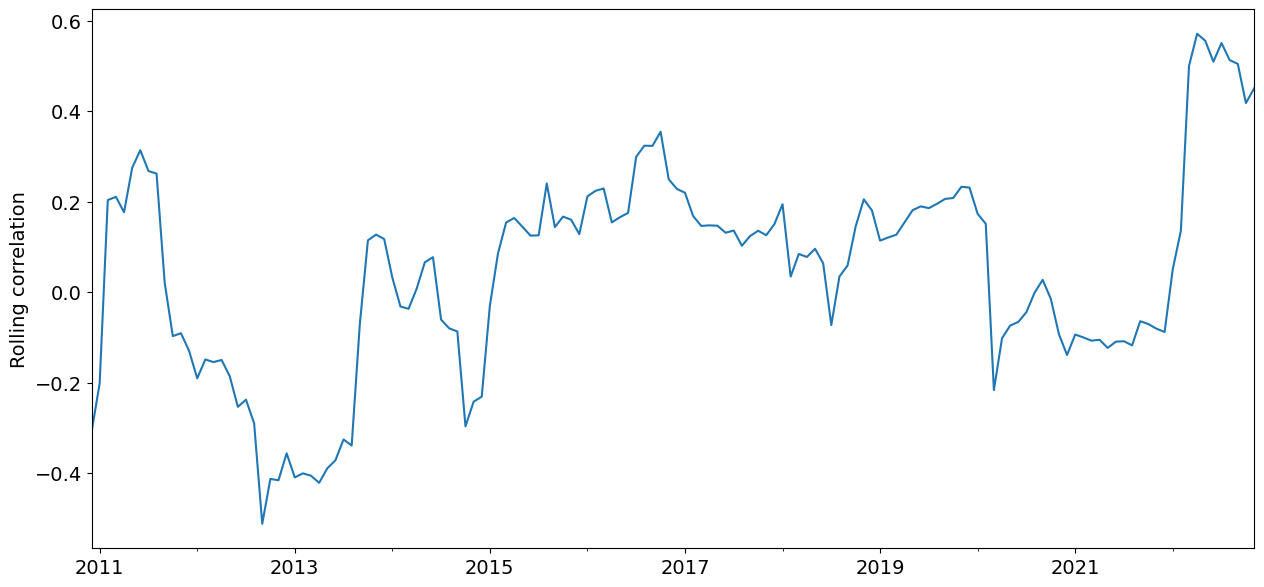}}
     \caption{\textbf{Rolling correlation between the returns on the long-short cyber risk portfolio and the market}}\bigskip
     \footnotesize{This figure depicts the rolling correlation between the monthly returns on the long-short portfolio built on the cyber risk score and the market portfolio. The rolling window is two years, and the period is from January 2011 to December 2022.}
\label{fig:rolling_correlation}
\end{figure}

\end{document}